\documentclass[11pt,a4paper]{article}

\usepackage{amsmath, amssymb}
\usepackage{braket}
\usepackage{cite}

\usepackage[top=25mm, bottom=25mm, left=25mm, right=25mm]{geometry}
\usepackage{titlesec}
\titlespacing*{\section}{0pt}{1ex}{-0.1ex}
\titlespacing*{\subsection}{0pt}{1ex}{-0.1ex}
\titlespacing*{\subsubsection}{0pt}{0ex}{-0.2ex}

\usepackage[colorlinks=true, linkcolor=blue, urlcolor=blue, citecolor=blue]{hyperref}

\usepackage{tikz}
\usetikzlibrary{arrows.meta}
\usetikzlibrary{decorations.pathmorphing}

\newcommand\raisemath[2]{\raisebox{#1}{$\displaystyle #2$} }

\newcommand{\be}{\begin{equation}}
\newcommand{\ee}{\end{equation}}
\newcommand{\bea}{\begin{eqnarray}}
\newcommand{\eea}{\end{eqnarray}}
\newcommand{\nn}{\nonumber}

\definecolor{desycyan}{rgb}{0.00,0.68,0.93}
\definecolor{desyorange}{rgb}{0.96,0.52,0.07}
\definecolor{desygray}{rgb}{0.47,0.47,0.47}
\definecolor{internationalorange}{rgb}{1.0, 0.31, 0.0}

\begin{document}
\thispagestyle{empty}
\begin{flushright}
\tt DESY 21-122
\end{flushright}

~
\vskip 30pt
\begin{center}
{\huge\sffamily\bfseries
The old conformal bootstrap revisited
}
\vskip 45pt

\textsc{Pedro Liendo}, \textsc{Zhengwen Liu} and \textsc{Junchen Rong}
\\[1.0 em]
{\it Deutsches Elektronen-Synchrotron DESY, Notkestra{\ss}e 85, 22607 Hamburg, Germany}
\\[0.5 em]
{Email}:\,\texttt{pedro.liendo@desy.de}, \texttt{zhengwen.liu@desy.de}, \texttt{junchen.rong@desy.de}
\end{center}

\vskip 20mm
\begin{quote}
\textbf{Abstract}:
The ``old'' conformal bootstrap was originally formulated by Migdal and Polyakov (MP) as a method for calculating conformal dimensions self-consistently. In this work we revisit the MP bootstrap and apply efficient multi-loop Feynman integral techniques in order to solve the corresponding equations. We obtain solutions at first order in the skeleton expansion for finite coupling and for integer values of the spacetime dimension. We focus in particular on $\phi^3$ theory in seven dimensions and the $O(N)$ vector models in three dimensions.
\end{quote}

%\newpage
%\tableofcontents

\newpage
\setcounter{page}{1}
\section{Introduction}
The Migdal-Polyakov (MP) bootstrap method, sometimes called the ``old'' bootstrap method, played an important role in understanding critical phenomena before the renormalization group was discovered \cite{patashinskii1964second,migdal1969diagra,polyakov1969microscopic,polyakov1970conformal,migdal1971conformal,parisi1971calculation,Mack:1972kq}. 
In fact, in his Nobel Prize lecture in 1982 Kenneth G.~Wilson said ``{\it if the 1971 renormalization group ideas had not been developed, the Migdal-Polyakov bootstrap would have been the most promising framework of its time for trying to further understand critical phenomena.}''\cite{wilsonnobel}.

Efforts to understand critical phenomena from conformal symmetry later lead to an alternative bootstrap approach, which is based on four-point functions in conformal field theories (CFTs) \cite{ferrara1973tensor,Polyakov:1974gs}. 
This is the formalism that became highly successful in the past decade and represents the modern conformal bootstrap program \cite{Rattazzi:2008pe}. In favorable situations, the numerical bootstrap method has been able to provide extremely precise predictions for scaling exponents  \cite{El-Showk:2014dwa,Kos:2014bka,Kos:2016ysd,Rong:2018okz,Atanasov:2018kqw,Chester:2019ifh}. For a recent review of the modern techniques see \cite{Poland:2018epd}. 

Unlike the modern bootstrap, the old MP bootstrap equations are based on three-point functions in CFTs, and can be expressed as series expansions in coupling constants using Feynman-like skeleton diagrams.
Each diagram can be written as a generalized multi-loop Feynman integral where the propagators can have arbitrary powers.
The evaluation of these integrals turned out to be a bottleneck for the MP bootstrap, the equations are technically challenging and were only solved in certain perturbative limits.
For example, in 1973, Mack discovered the first solution to the bootstrap equations of $\phi^3$ theory at leading order in 
$6+\epsilon$ dimensions  \cite{Mack:1973kaa}.
Another example is the large $N$ limit which has been used to solve $O(N)$ vector models \cite{Vasiliev:1981yc,vasil19821} and Gross-Neveu-Yukawa models \cite{Gracey:1990wi,Gracey:1993kx,Gracey:1993kc,Derkachov:1993uw}.

On a related line of research, the last decades have seen great progress in analytically and numerically evaluating multi-loop scattering amplitudes and Feynman integrals.
Any Feynman integral with arbitrary powers of propagators is equivalent to an integral over a set of parameters, such as the Feynman, Lee-Pomeransky \cite{Lee:2013hzt} and Mellin-Barnes \cite{Symanzik:1972wj,Smirnov:1999gc,Tausk:1999vh} representations, see \cite{Smirnov:2012gma} for a pedagogical review.
Once a representation is chosen, analytic continuation methods and advanced numerical integration techniques have proved to be a powerful method to evaluate dimensionally-regularized integrals.
Fortunately, these techniques are also applicable to the integrals that appear in the MP bootstrap equations.
We will solve the equations at first order in the skeleton expansion directly at finite coupling and for integer values of the spacetime dimension. This strategy was already attempted in the early days but was soon discarded due to the lack of proper technology. The main goal of this work is to revisit the MP bootstrap and apply modern multi-loop techniques to the evaluation of generalized Feynman integrals.

The paper is organized as follows. In Section \ref{phi3section} we discuss how to use multi-loop techniques to bootstrap the $\phi^3$ theory in seven dimensions. This model might be of pure academic interest, but the section also serves as a brief introduction to the Migdal-Polyakov bootstrap in a somehow simple setting. In Section \ref{ONmodels} we apply our techniques to the more physically relevant case of $O(N)$ vector models in three dimensions. In Section \ref{Discussionsection}, we discuss our results and propose possible future directions in order to generalize this work to higher orders.

\section{The $g\phi^3$ theory}\label{phi3section}

Let us start with the cubic $g\phi^3$ theory. This section serves as a brief introduction to the Migdal-Polyakov bootstrap, and as a test for our method in a setting where all the major steps can be easily explained.

The two point function (propagator) and the three-point one particle irreducible (1PI) vertex in $g \phi^3$ theory at the conformal fixed point are
\begin{align}
\label{2pt-function-phi3}
\langle\phi(x_1) \phi(x_2)\rangle &= \frac{1}{(x_{12}^2)^{\Delta}}\,,
\\[0.3 em]
\label{1PIvertex}
\langle\tilde{\phi}(x_1) \tilde{\phi}(x_2)  \tilde{\phi}(x_3)\rangle & = \frac{g}{(x_{12}^2)^{\frac{D-\Delta}{2}}(x_{23}^2)^{\frac{D-\Delta}{2}}(x_{31}^2)^{\frac{D-\Delta}{2}}}\,,
\end{align}
where $x_{ij} \equiv x_i {-} x_j$, $D$ is the dinension of spacetime, $\Delta$ is the conformal dimension of the field $\phi$, and the symbol $\tilde{\phi}$ stands for the shadow field of $\phi$, which has scaling dimension $\Delta_{\tilde{\phi}}=D-\Delta$. The three point correlation function is related to the three point 1PI vertex by (notice that $\langle\phi(x_1) \phi(x_2)\rangle$ is the full propagator after renormalization)
\begin{align}
\langle{\phi}(x_1) {\phi}(x_2) {\phi}(x_3)\rangle =\int d^D\! x_4 d^D\! x_5 d^D\! x_6\, &\langle\phi(x_1) \phi(x_4)\rangle\langle\phi(x_2) \phi(x_5)\rangle
\nonumber\\[-0.5 em]
\times&\langle\phi(x_3) \phi(x_6)\rangle \langle\tilde{\phi}(x_4) \tilde{\phi}(x_5)  \tilde{\phi}(x_6)\rangle\,.
\label{3ptintegral}
\end{align}
Under the conformal transformation $x \rightarrow x'$,
\begin{align}
d^D\!x ~\longrightarrow~ \bigg|\frac{\partial x'}{\partial x}\bigg|^{-D} d^D\!x', \quad
\phi(x) ~\longrightarrow~ \bigg|\frac{\partial x'}{\partial x}\bigg|^{\Delta} \phi(x')\,,
\quad
\tilde{\phi}(x) ~\longrightarrow~ \bigg|\frac{\partial x'}{\partial x}\bigg|^{D-\Delta} \tilde{\phi}(x')\,.
\end{align}
It is clear that the combination $\int d^D\!x \cdots \phi(x)\tilde{\phi}(x)\cdots$ is invariant under the conformal transformation.
The integral expression in \eqref{3ptintegral} therefore has the correct conformal covariance. 
To evaluate \eqref{3ptintegral}, we need the famous star-triangular formula introduced by Symanzik in \cite{Symanzik:1972wj},
\begin{align}\label{Gstartriangle}
\int d^D\! x_4 \frac{1}{(x^2_{14})^{\frac{\Delta_1}{2}}(x^2_{24})^{\frac{\Delta_2}{2}}(x^2_{34})^{\frac{\Delta_3}{2}}}=\frac{\kappa(\Delta_1,\Delta_2,\Delta_3)}{(x^2_{12})^{\frac{D-\Delta_3}{2}}(x^2_{23})^{\frac{D-\Delta_1}{2}}(x^2_{13})^{\frac{D-\Delta_2}{2}}}\,,
\end{align}
which is valid when $\Delta_1+\Delta_2+\Delta_3=2 D$. The function $\kappa$ is defined as
\begin{align}\label{}
\kappa(\Delta_1,\Delta_2,\Delta_3)=\pi^{D/2}\frac{\Gamma(\frac{D-\Delta_1}{2})\,\Gamma(\frac{D-\Delta_2}{2})\,\Gamma(\frac{D-\Delta_3}{2})}{\Gamma(\frac{\Delta_1}{2})\,\Gamma(\frac{\Delta_2}{2})\,\Gamma(\frac{\Delta_3}{2})}\,,
\end{align}
which is symmetric under permutation of its three arguments.
By iterating the star-triangular formula \eqref{3ptintegral} becomes 
\begin{align}\label{}
\langle\phi(x_1)\phi(x_2)\phi(x_3)\rangle = \kappa & (D{-}\Delta, D{-}\Delta, 2\Delta )\,
\kappa (\Delta, 2D{-}3\Delta, 2\Delta )\,
\kappa (2\Delta, D{-}\Delta, D{-}\Delta )
\nn\\
\times& \frac{g}{(x_{12}^2)^{\frac{\Delta}{2}}(x_{23}^2)^{\frac{\Delta}{2}}(x_{31}^2)^{\frac{\Delta}{2}}},
\end{align}
which is indeed a correct three-point function in a conformal field theory.

The three-point function with two scalars and a stress-energy tensor $\Braket{T_{\mu\nu} \phi \phi}$ is also fixed by conformal symmetry, i.e., 
\begin{align}\label{T-phi-phi}
\langle T_{\mu\nu}(x_1)\phi(x_2)\phi(x_3)\rangle=f_{\phi\phi T}\frac{Z_{\mu}Z_{\nu}-\frac{1}{D}\delta_{\mu\nu} Z_{\rho}Z^{\rho}}{(x_{12}^2)^{\frac{D-2}{2}}(x_{13}^2)^{\frac{D-2}{2}}(x_{23}^2)^{\frac{2\Delta-D+2}{2}}},
\end{align}
with
\begin{align}
Z^{\mu}=\frac{x_{12}^\mu}{x_{12}^2}-\frac{x_{13}^\mu}{x_{13}^2}.
\end{align}
The OPE coefficient $f_{\phi\phi T}$ is fixed by the corresponding Ward identity. 
The dilation operator is given by $\hat{D}=\int dS\, \hat{x}_{\mu}x_{\nu} T^{\mu\nu}$, with the integration over a sphere surrounding the origin\footnote{This is the dilation operator in radical quantization, for details, see \cite{Simmons-Duffin:2016gjk}.}. 
According to the conformal algebra, we need $\hat{D}\phi(0)=\Delta \phi(0)$. Using the OPE relation $ T_{\mu \nu }(x) \phi(0) = f_{\phi\phi T} \frac{1}{x^{D+2}} 
(x^{\mu } x^{\nu }- \frac{1}{D} \eta ^{\mu \nu } x^2 )\phi(0)+\ldots$, the OPE is fixed to be 
\be
f_{\phi\phi T}=-\frac{D\Delta_{}}{D-1}\frac{1}{S_{D-1}}\,,
\ee
where $S_{D-1}=\frac{\pi^{(D-1)/2}}{\Gamma((D+1)/2)}$ is the volume of the unit sphere $S^{D-1}$.  
The above three-point function satisfies the following differential form of the Ward identity \cite{Mack:1972kq}
\begin{align}\label{WardID}
\big\langle\partial_\nu T&^{\mu\nu}(x_1) \phi(x_2)\phi(x_3) \big\rangle
\nonumber\\
&= 2 \sqrt{\pi}\,\frac{ \Gamma \left(\frac{D-1}{2}\right)}{\Gamma \left(\frac{D}{2}\right)}
\Big(\delta^{D}(x_{12})\Braket{\partial^\mu \phi(x_2)\phi(x_3)} - {\Delta \over D} \big(\partial^\mu \delta^D(x_{12})\big)
\Braket{\phi(x_2)\phi(x_3)}
+ 2\leftrightarrow 3
\Big),
\end{align}
with $\partial_\mu \equiv \partial/\partial x_1^\mu$.

In \cite{polyakov1969microscopic} Polyakov realized that by analytically continuing the Euclidean field theories back to Lorentz signature, one could use the unitarity conditions to derive bootstrap equations for propagators. Later, Mack and Symanzik derived an alternative form of the the propagator bootstrap equations which involves the three point function $\langle T_{\mu\nu}\phi \phi \rangle$ \cite{Mack:1972kq}. This is the form of propagator bootstrap equations we will be using in this work, 
\begin{align}
\label{ppequations}
\begin{aligned}
\text{
\begin{tikzpicture}[scale=1]
\draw[-, color=desycyan, line width=0.8pt] (-0.8,0) -- (0.8,0);
\fill[color=white] (0,-0.1) circle (0pt);
\draw[-, color=internationalorange, line width=1.0pt] (-60:0.1) -- (120:0.1) -- (0,0) -- (240:0.1) -- (60:0.1);
\end{tikzpicture}
}
\end{aligned}
~{\color{desycyan}=}&~
\begin{aligned}
\text{
\begin{tikzpicture}[scale=1]
\draw[-, color=desycyan, line width=0.8pt] (0.6,0) arc (0:180:0.6 and 0.52);
\draw[-, color=desycyan, line width=0.8pt] (0.6,0) arc (0:-180:0.6 and 0.42);
\draw[-, color=desycyan, line width=0.8pt]  (-0.6,0) -- (-1.1,0);
\draw[-, color=desycyan, line width=0.8pt]  (0.6,0) -- (1.1,0);
\fill[color=desyorange] (-0.6,0) circle (3.0pt);
\fill[color=desyorange] (0.6,0) circle (3.0pt);
\draw[yshift=5.2mm, -, color=internationalorange, line width=1.0pt] (-60:0.1) -- (120:0.1) -- (0,0) -- (240:0.1) -- (60:0.1);
\end{tikzpicture}
}
\end{aligned}
~{\color{desycyan}+~}
\begin{aligned}
\text{
\begin{tikzpicture}[scale=1]
\draw[-, color=desycyan, line width=0.8pt] (1.2,0) arc (0:180:0.8 and 0.55);
\draw[-, color=desycyan, line width=0.8pt] (0.4,0) arc (0:-180:0.8 and 0.5);
\draw[-, color=desycyan, line width=0.8pt] (-0.4,0) -- (0.4,0);
\draw[-, color=desycyan, line width=0.8pt] (-1.2,0) -- (-0.4,0);
\draw[-, color=desycyan, line width=0.8pt] (1.2,0) -- (0.4,0);
\draw[-, color=desycyan, line width=0.8pt]  (-1.2,0) -- (-1.7,0);
\draw[-, color=desycyan, line width=0.8pt]  (1.2,0) -- (1.7,0);
\fill[color=desyorange] (-0.4,0) circle (3.0pt);
\fill[color=desyorange] (0.4,0) circle (3.0pt);
\fill[color=desyorange] (-1.2,0) circle (3.0pt);
\fill[color=desyorange] (1.2,0) circle (3.0pt);
\draw[-, color=internationalorange, line width=1.0pt] (-60:0.1) -- (120:0.1) -- (0,0) -- (240:0.1) -- (60:0.1);
\end{tikzpicture}
}
\end{aligned}
~{\color{desycyan}+\cdots}
\end{align}
Here
\tikz{
\draw[-, color=desycyan, line width=0.6pt] (-0.5,0) -- (0.5,0);
\draw[-, color=internationalorange, line width=0.8pt] (-60:0.1) -- (120:0.1) -- (0,0) -- (240:0.1) -- (60:0.1);
} denotes the three point correlation function 
$\Braket{T_{\mu\nu} \phi \phi}$ defined in \eqref{T-phi-phi}, and the orange dot (connected to three legs) denotes the 1PI vertex \eqref{1PIvertex}, each internal line that connects two vertices denotes the two-point function \eqref{2pt-function-phi3}.
The alternative forms of the propagator bootstrap equation from unitarity conditions can be found in \cite{symanzik1961green,migdal1969diagra,polyakov1969microscopic,parisi1971calculation,Parisi:1972zm}. 
For more details, see for example \cite{Mack:1973kaa,Fradkin:1978pp,Fradkin:1996is}.
Similarly, the vertex bootstrap equation was proposed by Migdal in \cite{migdal1971conformal},
\begin{align}\label{bootstrapeqns}
\begin{aligned}
\text{
\begin{tikzpicture}[scale=1]
\draw[-, color=desycyan, line width=0.8pt] (0,0) -- (90:0.4);
\draw[-, color=desycyan, line width=0.8pt] (0,0) -- (220:0.5);
\draw[-, color=desycyan, line width=0.8pt] (0,0) -- (320:0.5);
\fill[color=desyorange] (0,0) circle (3.0pt);
\end{tikzpicture}
}
\end{aligned}
~~{\color{desycyan}=}&~
\begin{aligned}
\text{
\begin{tikzpicture}[scale=1]
\draw[-, color=desycyan, line width=0.8pt] (-30:0.8) -- (90:0.8) -- (210:0.8);
\draw[-, color=desycyan, line width=0.8pt] (210:0.8) -- (-30:0.8);
\draw[-, color=desycyan, line width=0.8pt] (90:0.8) -- (90:1.1);
\draw[-, color=desycyan, line width=0.8pt] (-30:0.8) -- (320:1.1);
\draw[-, color=desycyan, line width=0.8pt] (210:0.8) -- (220:1.1);
\fill[color=desyorange] (-30:0.8) circle (3.0pt);
\fill[color=desyorange] (90:0.8) circle (3.0pt);
\fill[color=desyorange] (210:0.8) circle (3.0pt);
\end{tikzpicture}
}
\end{aligned}
~{\color{desycyan}+}~
\begin{aligned}
\text{
\begin{tikzpicture}[scale=1]
\draw[-, color=desycyan, line width=0.8pt]  (35:0.8) -- (90:1.2) -- (145:0.8);
\draw[-, color=desycyan, line width=0.8pt] (35:0.8) -- (-35:0.8);
\draw[-, color=desycyan, line width=0.8pt] (145:0.8) -- (-145:0.8);
\draw[-, color=desycyan, line width=0.8pt] (145:0.8) -- (-35:0.8);
\draw[-, color=desycyan, line width=0.8pt] (-145:0.8) -- (-145:0.1);
\draw[-, color=desycyan, line width=0.8pt] (35:0.8) -- (35:0.1);
\draw[-, color=desycyan, line width=1.0pt] (90:1.2) -- (90:1.5);
\draw[-, color=desycyan, line width=0.8pt] (-145:0.8) -- (215:1.1);
\draw[-, color=desycyan, line width=0.8pt] (-35:0.8) -- (325:1.1);
\fill[color=desyorange] (90:1.2) circle (3.0pt);
\fill[color=desyorange] (35:0.8) circle (3.0pt);
\fill[color=desyorange] (-35:0.8) circle (3.0pt);
\fill[color=desyorange] (145:0.8) circle (3.0pt);
\fill[color=desyorange] (-145:0.8) circle (3.0pt);
\end{tikzpicture}
}
\end{aligned}
~{\color{desycyan}+ \cdots}
\end{align}
Physically, the above bootstrap equations mean that the corresponding solutions describe correlators that are invariant under renormalization.
For each bootstrap equation, \eqref{ppequations} or \eqref{bootstrapeqns}, the right hand side is a skeleton expansion.
It is also a series in $g$ where the number of 1PI vertices counts the power of $g$. At a given order in $g$, all inequivalent skeleton diagrams are included.
It is crucial that diagrams like
\begin{align}\label{}
\begin{aligned}
\text{
\begin{tikzpicture}[scale=1]
\draw[-, color=desycyan, line width=0.8pt] (90:1.2) -- (35:0.8) -- (145:0.8) -- (90:1.2);
\draw[-, color=desycyan, line width=0.8pt] (-35:0.8) -- (-145:0.8);
\draw[-, color=desycyan, line width=0.8pt] (35:0.8) -- (-35:0.8);
\draw[-, color=desycyan, line width=0.8pt] (145:0.8) -- (-145:0.8);
\draw[-, color=desycyan, line width=0.8pt] (90:1.2) -- (90:1.5);
\draw[-, color=desycyan, line width=0.8pt] (-145:0.8) -- (215:1.1);
\draw[-, color=desycyan, line width=0.8pt] (-35:0.8) -- (325:1.1);
\fill[color=desyorange] (90:1.2) circle (3.0pt);
\fill[color=desyorange] (35:0.8) circle (3.0pt);
\fill[color=desyorange] (-35:0.8) circle (3.0pt);
\fill[color=desyorange] (145:0.8) circle (3.0pt);
\fill[color=desyorange] (-145:0.8) circle (3.0pt);
\draw[dotted,line width=0.7pt] (0,0.7) ellipse (28pt and 19pt);
\end{tikzpicture}
}
\end{aligned}
\end{align}
should not be included, as the circled part is a renormalization of the 1PI vertex, which is already included in the fully dressed 1PI vertex $\langle\tilde{\phi} \tilde{\phi} \tilde{\phi}\rangle$.
Because of this property, at a fixed order, the skeleton expansion contains much less diagrams than the usual Feynman diagram expansion.

In addition to the three external coordinates $\{x_1, x_2, x_3\}$, in the two ends of each internal line, there are two undetermined (loop) coordinates, which need to be integrated.
An important property is that in every diagram, $\phi(x)$, $\tilde{\phi}(x)$ and $\int dx^D$ always appear together. This guarantees that the diagram has the correct conformal transformation to be consistent with the left-hand side of the equation.
Using various building blocks, including \eqref{2pt-function-phi3}, \eqref{1PIvertex} and \eqref{T-phi-phi}, all skeleton graphs can be translated into generalized Feynman integrals, where the powers of propagators are the linear combinations of conformal weights and the spacetime dimension.
The two equations \eqref{ppequations} and \eqref{bootstrapeqns} have two unknowns, and in principle can be used to solve for $\{\Delta, g^2\}$. This is the point where the classic papers stop, solving these equations is challenging due to the complexity of the integrals involved.

In the following we describe a framework to calculate generalized Feynman integrals in order to solve the MP bootstrap equations.
In this work we focus on the leading order (LO) in the skeleton expansion, which already captures all the key steps of our method.
At LO, as shown in \eqref{ppequations} and \eqref{bootstrapeqns}, both the propagator and vertex bootstrap equations have only one diagram.
Let us start with the vertex equation \eqref{bootstrapeqns}.
Using the two-point function \eqref{2pt-function-phi3} and the 1PI vertex \eqref{1PIvertex}, the first diagram on the right-hand side of \eqref{bootstrapeqns} can be written as
\begin{align}\label{vertexintegral}
I_1(x_1,x_2,x_3) = \Bigg(\prod_{i=4}^{9}\int d^D\! x_i \!\Bigg) & \langle\tilde{\phi}(x_1) \tilde{\phi}(x_4)  \tilde{\phi}(x_5)\rangle\langle\tilde{\phi}(x_2) \tilde{\phi}(x_6)  \tilde{\phi}(x_7)\rangle \langle\tilde{\phi}(x_3) \tilde{\phi}(x_8)  \tilde{\phi}(x_9)\rangle
\nonumber\\[-0.7 em] 
&\times \langle\phi(x_4) \phi(x_6)\rangle\langle\phi(x_5) \phi(x_9)\rangle\langle\phi(x_7) \phi(x_8)\rangle
\nonumber\\[0.2 em] 
= \Bigg(\prod_{i=4}^{9}\int d^D\! x_i \!\Bigg) &
{g^3 \over \big(x_{14}^2 x_{15}^2 x_{45}^2 x_{26}^2 x_{27}^2 x_{67}^2 x_{38}^2 x_{39}^2 x_{89}^2\big)^{(D-\Delta)/2}
\big(x_{56}^2 x_{78}^2 x_{49}^2\big)^{\Delta}}.
\end{align}
This is a six-loop generalized Feynman integral.
Luckily, we can easily evaluate the integrals over three of the internal coordinates using the star-triangle formula \eqref{Gstartriangle}. The remaining integrals are highly non-trivial, and our strategy is to employ modern multi-loop techniques.
More explicitly, we first derive a parametric representation for each integral in our computation. There are many options available such as MB, Feynman and Lee-Pomeransky representations (see Appendix \ref{sec-app-FI} for a short summary).
Notice that these parametric integrals are, in general, not well-defined in some  physical regions. Usually, they need to be analytically continued to the region of physical interest.
Many methods have been developed to perform the analytic continuation systematically, such as the MB technique \cite{Czakon:2005rk,Tausk:1999vh} and the sector decomposition method \cite{Hepp:1966eg,Speer:1975dc,Binoth:2000ps,Binoth:2003ak}.
Finally, various efficient numerical integration methods and tools can be used to numerically evaluate resulting integrals, for example \texttt{Cuba} \cite{Hahn:2004fe}.

Below we compute the integral in \eqref{vertexintegral} in detail to show our method more explicitly.
We first translate the integral into an MB integral representation.
To proceed, let us introduce a nice graphic representation for generalized Feynman integrals\,--\,Migdal diagrams.
In these diagrams, each line connecting two coordinates represents a propagator, while the weight on the edge denotes the power of the propagator.
In this way, the integral in \eqref{vertexintegral} can be represented by the following left diagram.
\begin{align}\label{Migdal1}
\begin{aligned}
\text{
\begin{tikzpicture}[scale=1]
\coordinate [] (x1) at (90:2.77128);  % 2Sqrt[3] * 0.8
\coordinate [] (x2) at (-150:2.77128);
\coordinate [] (x3) at (-30:2.77128);
\coordinate [] (x4) at (60:1.6);
\coordinate [] (x9) at (0:1.6);
\coordinate [] (x8) at (-60:1.6);
\coordinate [] (x7) at (-120:1.6);
\coordinate [] (x6) at (180:1.6);
\coordinate [] (x5) at (120:1.6);
\draw[-, color=desycyan, line width=0.8pt] (x1) -- (x2) -- (x3) -- (x1) -- (x2);
\draw[-, color=desycyan, line width=0.8pt] (x4) -- (x5);
\draw[-, color=desycyan, line width=0.8pt] (x6) -- (x7);
\draw[-, color=desycyan, line width=0.8pt] (x8) -- (x9);
\fill[color=black] (x1) circle (0) node[yshift=1.5mm] {\large $x_1$};
\fill[color=black] (x2) circle (0) node[xshift=-2mm, yshift=-1.3mm] {\large $x_2$};
\fill[color=black] (x3) circle (0) node[xshift=2.2mm, yshift=-1.3mm] {\large $x_3$};
\fill[color=black] (x4) circle (0) node[right] {\large $x_4$};
\fill[color=black] (x5) circle (0) node[left] {\large $x_5$};
\fill[color=black] (x9) circle (0) node[right] {\large $x_9$};
\fill[color=black] (x6) circle (0) node[left] {\large $x_6$};
\fill[color=black] (x7) circle (0) node[below] {\large $x_7$};
\fill[color=black] (x8) circle (0) node[below] {\large $x_8$};
\fill[color=black] (80:2.15) circle (0) node[right] {\small$D{-}\Delta$};
\fill[color=black] (100:2.15) circle (0) node[left] {\small$D{-}\Delta$};
\fill[color=black] (30:1.4) circle (0) node[right] {\small$2\Delta$};
\fill[color=black] (150:1.4) circle (0) node[left] {\small$2\Delta$};
\fill[color=black] (-17:2.0) circle (0) node[right] {\small$D-\Delta$};
\fill[color=black] (197:2.0) circle (0) node[left] {\small$D{-}\Delta$};
\fill[color=black] (219:2.12) circle (0) node[below] {\small$D{-}\Delta$};
\fill[color=black] (-40:2.12) circle (0) node[below] {\small$D{-}\Delta$};
\fill[color=black] (-90:1.35) circle (0) node[below] {\small$2\Delta$};
\fill[color=black] (90:1.2) circle (0) node[] {\small$D{-}\Delta$};
\fill[color=black] (210:1.41) circle (0) node[right] {\small$D{-}\Delta$};
\fill[color=black] (-30:1.41) circle (0) node[left] {\small$D{-}\Delta$};
\end{tikzpicture}
}
\end{aligned}
\quad
%%%%
%%%%
\begin{aligned}
\text{
\begin{tikzpicture}[scale=1]
\coordinate [] (x1) at (90:2.77128);  % 2Sqrt[3] * 0.8
\coordinate [] (x2) at (-150:2.77128);
\coordinate [] (x3) at (-30:2.77128);
\coordinate [] (x4) at (60:1.6);
\coordinate [] (x8) at (-60:1.6);
\coordinate [] (x6) at (180:1.6);
\draw[-, color=desycyan, line width=0.8pt] (x1) -- (x2) -- (x3) -- (x1) -- (x2);
\draw[-, color=desycyan, line width=0.8pt] (x4) -- (x6) -- (x8) -- (x4);
\fill[color=black] (x1) circle (0) node[yshift=1.5mm] {\large $x_1$};
\fill[color=black] (x2) circle (0) node[xshift=-2mm, yshift=-1.3mm] {\large $x_2$};
\fill[color=black] (x3) circle (0) node[xshift=2.2mm, yshift=-1.3mm] {\large $x_3$};
\fill[color=black] (x4) circle (0) node[right] {\large $x_4$};
\fill[color=black] (x6) circle (0) node[left] {\large $x_6$};
\fill[color=black] (x8) circle (0) node[below] {\large $x_8$};
\fill[color=black] (80:2.15) circle (0) node[right] {\small$2D{-}3\Delta$};
\fill[color=black] (130:1.5) circle (0) node[left] {\small$\Delta$};
\fill[color=black] (0:1.6) circle (0) node[right] {\small$\Delta$};
\fill[color=black] (235:1.15) circle (0) node[left] {\small$2D{-}3\Delta$};
\fill[color=black] (-30:1.75) circle (0) node[below] {\small$2\Delta{-}3\Delta$};
\fill[color=black] (-125:1.65) circle (0) node[below] {\small$\Delta$};
\fill[color=black] (0:0.6) circle (0) node[] {\small$\Delta$};
\fill[color=black] (-105:0.5) circle (0) node[] {\small$\Delta$};
\fill[color=black] (120:0.65) circle (0) node[] {\small$\Delta$};
\end{tikzpicture}
}
\end{aligned}
\end{align}
As mentioned previously, we can easily integrate out three of six internal coordinates using the star-triangular formula \eqref{Gstartriangle}, and obtain
\begin{align}
I_1=\kappa^3(D{-}\Delta, D{-}\Delta, 2\Delta) \times I_2,
\end{align}
where $I_2$ is represented by the right diagram in \eqref{Migdal1} (it is straightforward to write down a mathematical expression from the Migdal diagram).
We observed that all three loop integrals are four-star integrals, i.e.\,each node associated with an internal coordinate connects 4 lines.
Furthermore, for each integration coordinate, the sum of the powers of all four related propagators is equal to $2D$, which is consistent with the property that the integral has a correct conformal weight as the left-hand side of the bootstrap equation.
Because of this property, we can successively use Symanzik's generalization of the star-triangular formula \eqref{symanzik-mb} to write the integral as an MB representation.
Schematically,
\begin{align}\label{Migdal2}
\begin{aligned}
\text{
\begin{tikzpicture}[scale=1]
\coordinate [] (x1) at (90:1.73205);  % 2Sqrt[3] * 0.5
\coordinate [] (x2) at (-150:1.73205);
\coordinate [] (x3) at (-30:1.73205);
\coordinate [] (x8) at (-90:0.866025);
\coordinate [] (x6) at (150:0.866025);
\draw[-, color=desycyan, line width=0.8pt] (x1) -- (x2) -- (x3) -- (x1) -- (x2);
\draw[-, color=desycyan, line width=0.8pt] (x1) -- (x8);
\draw[-, color=desycyan, line width=0.8pt] (x6) -- (x8);
\draw[-, color=desycyan, line width=0.8pt] (x6) -- (150:0.1);
\draw[-, color=desycyan, line width=0.8pt] (x3) -- (-30:0.1);
\fill[color=black] (x1) circle (0) node[yshift=1.5mm] {\large $x_1$};
\fill[color=black] (x2) circle (0) node[xshift=-2mm, yshift=-1.3mm] {\large $x_2$};
\fill[color=black] (x3) circle (0) node[xshift=2.2mm, yshift=-1.3mm] {\large $x_3$};
\fill[color=black] (x6) circle (0) node[left] {\large $x_6$};
\fill[color=black] (x8) circle (0) node[below] {\large $x_8$};
\end{tikzpicture}
}
\end{aligned}
\quad
%%%%
%%%%
\begin{aligned}
\text{
\begin{tikzpicture}[scale=1]
\coordinate [] (x1) at (90:1.73205);  % 2Sqrt[3] * 0.5
\coordinate [] (x2) at (-150:1.73205);
\coordinate [] (x3) at (-30:1.73205);
\coordinate [] (x8) at (-90:0.866025);
\coordinate [] (x6) at (150:0.866025);
\draw[-, color=desycyan, line width=0.8pt] (x1) -- (x2) -- (x3) -- (x1) -- (x2);
\draw[-, color=desycyan, line width=0.8pt] (x1) -- (x8);
\fill[color=black] (x1) circle (0) node[yshift=1.5mm] {\large $x_1$};
\fill[color=black] (x2) circle (0) node[xshift=-2mm, yshift=-1.3mm] {\large $x_2$};
\fill[color=black] (x3) circle (0) node[xshift=2.2mm, yshift=-1.3mm] {\large $x_3$};
\fill[color=black] (x8) circle (0) node[below] {\large $x_8$};
\end{tikzpicture}
}
\end{aligned}
\quad
%%%%
%%%%
\begin{aligned}
\text{
\begin{tikzpicture}[scale=1]
\coordinate [] (x1) at (90:1.73205);  % 2Sqrt[3] * 0.5
\coordinate [] (x2) at (-150:1.73205);
\coordinate [] (x3) at (-30:1.73205);
\coordinate [] (x8) at (-90:0.866025);
\coordinate [] (x6) at (150:0.866025);
\draw[-, color=desycyan, line width=0.8pt] (x1) -- (x2) -- (x3) -- (x1) -- (x2);
\fill[color=black] (x1) circle (0) node[yshift=1.5mm] {\large $x_1$};
\fill[color=black] (x2) circle (0) node[xshift=-2mm, yshift=-1.3mm] {\large $x_2$};
\fill[color=black] (x3) circle (0) node[xshift=2.2mm, yshift=-1.3mm] {\large $x_3$};
\end{tikzpicture}
}
\end{aligned}
\end{align}
From the second diagram in \eqref{Migdal1}, we obtain the first one above after `integrating' over $x_4$ using the Symanzik formula \eqref{symanzik-mb}, similarly we can iteratively integrate over $x_6$ and $x_8$ to obtain the final MB representation:
\begin{align}\label{}
I_1= \hat{\mathcal{I}}(D,\Delta) \times \frac{g^3 \kappa^3(D{-}\Delta, D{-}\Delta, 2\Delta)}{(x_{12}^2)^{\frac{D-\Delta}{2}}(x_{23}^2)^{\frac{D-\Delta}{2}}(x_{31}^2)^{\frac{D-\Delta}{2}}},
\end{align}
with
\begin{align}\label{MBintegral1loop}
\hat{\mathcal{I}}(D,\Delta)
=\pi^{D}\int_{-i \infty}^{i\infty}  & ds_1 ds_2 dt_1 dt_2
\frac{\kappa \left(\frac{D}{2}+s_2+2 t_2, \frac{D}{2} + \frac{\Delta}{2} - s_1 - s_2 - 2t_1 - 2t_2, D-\frac{\Delta }{2} + s_1+2 t_1\right)}{\Gamma^3(\frac{\Delta }{2})\,\Gamma^2(D-\frac{3 \Delta }{2})\, \Gamma(\frac{D}{3}+s_1)\, \Gamma(\Delta-\frac{D}{6} -s_1-s_2)}
\nonumber\\
&\times \Gamma\big(\tfrac{D}{3}-\tfrac{\Delta }{2}+s_1\big)
\Gamma\big(\tfrac{\Delta }{2}-\tfrac{D}{6}+s_1\big)
\Gamma\big(\tfrac{D}{3}-\tfrac{\Delta }{2}-s_1-s_2\big)
\Gamma\big(\tfrac{\Delta }{2}-\tfrac{D}{6}-s_1-s_2\big)
\nonumber\\
&\times \Gamma\big(\tfrac{D}{3}-\tfrac{\Delta }{2}+s_2\big)
\Gamma\big(\tfrac{3 \Delta }{4}-\tfrac{D}{3}-\tfrac{1}{2}s_1+t_1\big)
\Gamma\big(\tfrac{D}{2}-\tfrac{3 \Delta }{4}+\tfrac{1}{2}s_1+t_1\big)
\nonumber\\
&\times \Gamma\big(\tfrac{\Delta}{4}  - \tfrac{D}{12} + \tfrac{1}{2}s_1+ \tfrac{1}{2}s_2-t_1-t_2\big)
\Gamma\big(\tfrac{\Delta }{2} -\tfrac{D}{12}-\tfrac{1}{2}s_2+t_2\big)\,
\nonumber\\
&\times
\Gamma\big(\tfrac{D}{4} - \tfrac{\Delta}{4} - \tfrac{1}{2}s_1 - \tfrac{1}{2}s_2 - t_1 - t_2\big)
\Gamma\big(\tfrac{D}{4} - \tfrac{\Delta}{2} + \tfrac{1}{2}s_2 + t_2\big) .
\end{align}
Here the contour is defined such that all the $\Gamma$'s in the numerator of the integrand (including the $\Gamma$'s in $\kappa$) to have positive real parts (which is required by Symmanzik's generalization of star-triangular formula).
This is only possible if $$\frac{D}{3}\leq \Delta\leq\frac{D}{2}.$$ 
In fact, in such a region, the integral in \eqref{vertexintegral} or\eqref{MBintegral1loop} is free of UV and IR divergence. 
The existence of such a region was conjectured in the original work of Migdal \cite{migdal1971conformal}, and was proven for Yukawa theory in  \cite{Mack:1973mq}, see also Section 3 of \cite{Mack:1973kaa}. 
It was also proven in the earlier days that the generalized Feynman integral is a meromorphic function of $\Delta$ \cite{Speer1969gfa}, which is an analytic continuation of the integral \eqref{vertexintegral}.
The MB representation of Migdal diagrams was introduced by Symmanzik in \cite{Symanzik:1972wj}. To evaluate this integral, however, another step is necessary, namely analytical continuation. 
Suppose the theory we study lives beyond the convergent domain, such an analytic continuation is necessary.  
One can define the analytic continuation of integrals by changing the integration contour. 
For example, the following integral\footnote{This kind of integral typically appears in instanton calculation.} 
 \begin{align}\label{}
 K(g)=\int_{\mathcal C}  e^{-\frac{1}{4} g \phi^4-\frac{1}{2}\phi^2} d\phi,
 \end{align}
with $\operatorname{Re}(g)>0$ is defined by choosing the integration contour $\mathcal C$ to be along the real axis. To continue $K(g)$ to $\operatorname{Re}(g)<0$, however, the integration contour $\mathcal C$ need to approach $\operatorname{arg}(\phi)=\pm\frac{\pi}{4}$ at large $|\phi|$, see for example, \cite{McKane:1978md}.   
Our case is a bit more complicated, when $\Delta< D/3$, it is not possible to pick a contour to guarantee that all the arguments of the  $\Gamma$ function in the numerator of the integrand in \eqref{MBintegral1loop} to have positive real parts. 
This is because $\mathcal{I}(D,\Delta)$, as a function of $\Delta$, has a pole at $\Delta= D/3$.

To proceed, let us show how to perform the analytic continuation for MB integrals in detail.
For simplicity, let us look at the following example \cite{Tausk:1999vh,Smirnov:2012gma}
\begin{align}\label{}
\int_{-i\infty}^{i\infty} dz\,\Gamma(-z)\,\Gamma(\lambda+z).
\end{align}
We pick the integration contour to be parallel to the imaginary axis. 
The formula is only valid if the integration contour separates the poles of $\Gamma(-z)$ and $\Gamma(\lambda+z)$.
The separation of the poles is only possible when $\operatorname{Re}(\lambda)>0$.
To analytically continue to the region $-1< \operatorname{Re}(\lambda) <0$, one can pick up the pole at $z=-\lambda$ by deforming the integration contour. 
To be explicit,
\begin{align}\label{}
{2\pi i}\,\Gamma(\lambda)
+ \int_{\mathcal{C}}dz\, \Gamma(-z)\,\Gamma(\lambda+z),
\end{align}
where the deformed contour $\mathcal{C}$ is now along the imaginary axis and intersect with the real axis in the region ($-\lambda-1,0$). 
The first term contains a simple pole at $\lambda=0$, while the second term is regular at $\lambda=0$. 
For given values of $\lambda$, now one can directly evaluate the integral analytically or numerically.
For more complicated integrals, like our integral in \eqref{MBintegral1loop}, similar algorithms have been proposed and implemented in publicly available programs, such as \texttt{MB.m}\footnote{The online version need the library \texttt{cernlib.lib} to work with Gamma functions \cite{Czakon:2005rk}.
}.

We used \texttt{MB.m} to analytically continue our integral \eqref{MBintegral1loop} to various physical regions we are interested in.
Then we numerically computed the integrals involved using the numerical integration library \texttt{Cuba} \cite{Hahn:2004fe}.
We present the results at the end of this section.

Now let us turn to the propagator bootstrap equation \eqref{ppequations}.
As mentioned previously, at LO in $g$, there is only one diagram and can be written as
\begin{align}\label{pp-eq-lo-integral}
\Bigg(\prod_{i=4}^{9}\int d^D\! x_i \!\Bigg)
&
\big\langle \phi (x_6) \phi(x_7) \big\rangle
\big\langle \phi(x_8) \phi(x_2) \big\rangle
\big\langle \phi(x_9) \phi(3) \big\rangle
\\[-0.5 em]
&\times
\big\langle \tilde{\phi}(x_4) \tilde{\phi}(x_6) \tilde{\phi}(x_8) \big\rangle
\big\langle \tilde{\phi}(x_5) \tilde{\phi}(x_7) \tilde{\phi}(x_9) \big\rangle
\big\langle \phi(x_4) \phi(x_5) T_{\mu\nu }(x_1) \big\rangle.
\nonumber
\end{align}
Following the method we described previously, we can derive an MB representation for the formula above, i.e.
\begin{align}
g^2 \mathcal{X}(D,\Delta)\, \big\langle T_{\mu\nu}(x_1) \phi(x_2) \phi(x_3)\big\rangle,
\end{align}
where $\mathcal{X}(D,\Delta)$ is a multiple MB integral.
Of course, we can resolve the integral with \texttt{MB.m} and numerically evaluate it using \texttt{Cuba} in various physical regions.
Remarkably, we can analytically compute the integral in \eqref{pp-eq-lo-integral} using a trick introduced in \cite{Fradkin:1996is} based on Ward identity \eqref{WardID}.
The final result is
\begin{align}\label{prop-eq-int-res}
\mathcal{X}(D,\Delta) =\,& \frac{\pi^{3 D} (\Delta -1)\, \Gamma(\Delta -1)\, \Gamma^3(\frac{\Delta }{2})}{2\Delta D (D-\Delta -1) (D-\Delta ) \Gamma(\frac{D}{2})\, \Gamma^5(\Delta )\, \Gamma(D-\frac{3 \Delta }{2})}
\nn\\
&\times \frac{\Gamma^5(\frac{D}{2}-\Delta)\, \Gamma(\Delta - \frac{D}{2} +1)\, \Gamma(\frac{3 \Delta }{2}-\frac{D}{2})}
{\Gamma(\frac{D}{2}-\Delta +1)\,\Gamma(D-\Delta -1)\, \Gamma^3(\frac{D-\Delta }{2})}
\\
&\times\Big(
D^2\Delta\, \psi\big(\tfrac{\Delta }{2}\big)
- D^2\Delta\, \psi\big(\tfrac{3 \Delta }{2} {-} \tfrac{D}{2}\big)
-2 D^2 - 8 \Delta ^2
- D\Delta^2\, \psi\big(\tfrac{\Delta }{2}\big)
+8\Delta D
\nn\\
&\qquad +D \Delta^2 \,\psi\big(\tfrac{3 \Delta }{2} {-} \tfrac{D}{2}\big)
+\Delta  D (D-\Delta )\, \psi\big(\tfrac{D-\Delta }{2}\big)
+\Delta  D (\Delta -D)\, \psi\big(D {-} \tfrac{3 \Delta }{2}\big)\Big),
\nonumber
\end{align}
where $\psi$ is the digamma function.
The calculation is lengthy and the details will be given in Appendix \ref{Stresstensoreqn}.
We also numerically checked this integral for many values for $D$ and $\Delta$ using \texttt{MB.m} and \texttt{Cuba}.

In summary, the vertex and propagator bootstrap equations to leading order can be written as 
\begin{align}
1 &= g^2\, \mathcal{I}(D,\Delta_{})+\cdots,
\\
1 &= g^2 \mathcal{X}(D,\Delta_{})+\cdots,
\end{align}
where $\mathcal{I}(D,\Delta) = \kappa^3(D{-}\Delta, D{-}\Delta, 2\Delta) \times \hat{\mathcal{I}}(D,\Delta)$ with $\hat{\mathcal{I}}$ defined in \eqref{MBintegral1loop}.
The two bootstrap equations have two unknowns $\{g^2, \Delta\}$ and are highly non-linear.
As we will show later, $ \mathcal{I}(D,\Delta_{})$ has a pole at $\Delta_{}=\frac{D}{3}$, while $ \mathcal{X}(D,\Delta_{})$ has a pole at $\Delta_{}=\frac{D-2}{2}$.
At $D=6$, the two poles coincide.
Using this property, Mack solves the bootstrap equations in $6+\epsilon$ dimensions\cite{Mack:1973kaa}.
Setting $D=6+\epsilon$ and $\Delta_{}=\frac{D-2}{2}+\alpha \epsilon$, we get
\begin{align}
\mathcal{I}(D,\Delta_{}) &= g^2 \frac{8 \pi ^{18}}{(6 \alpha -1)^3 \epsilon ^3}+\cdots \nn\\
\mathcal{X}(D,\Delta_{}) &= g^2 \frac{\pi ^{18}}{3 \alpha  (6 \alpha +1)^2 \epsilon ^3}+\cdots,
\end{align}
which gives us a solution with $\alpha=\frac{1}{18}$.

\begin{figure}[ht]
\begin{center}
\hspace{4pt}\includegraphics[width=10cm]{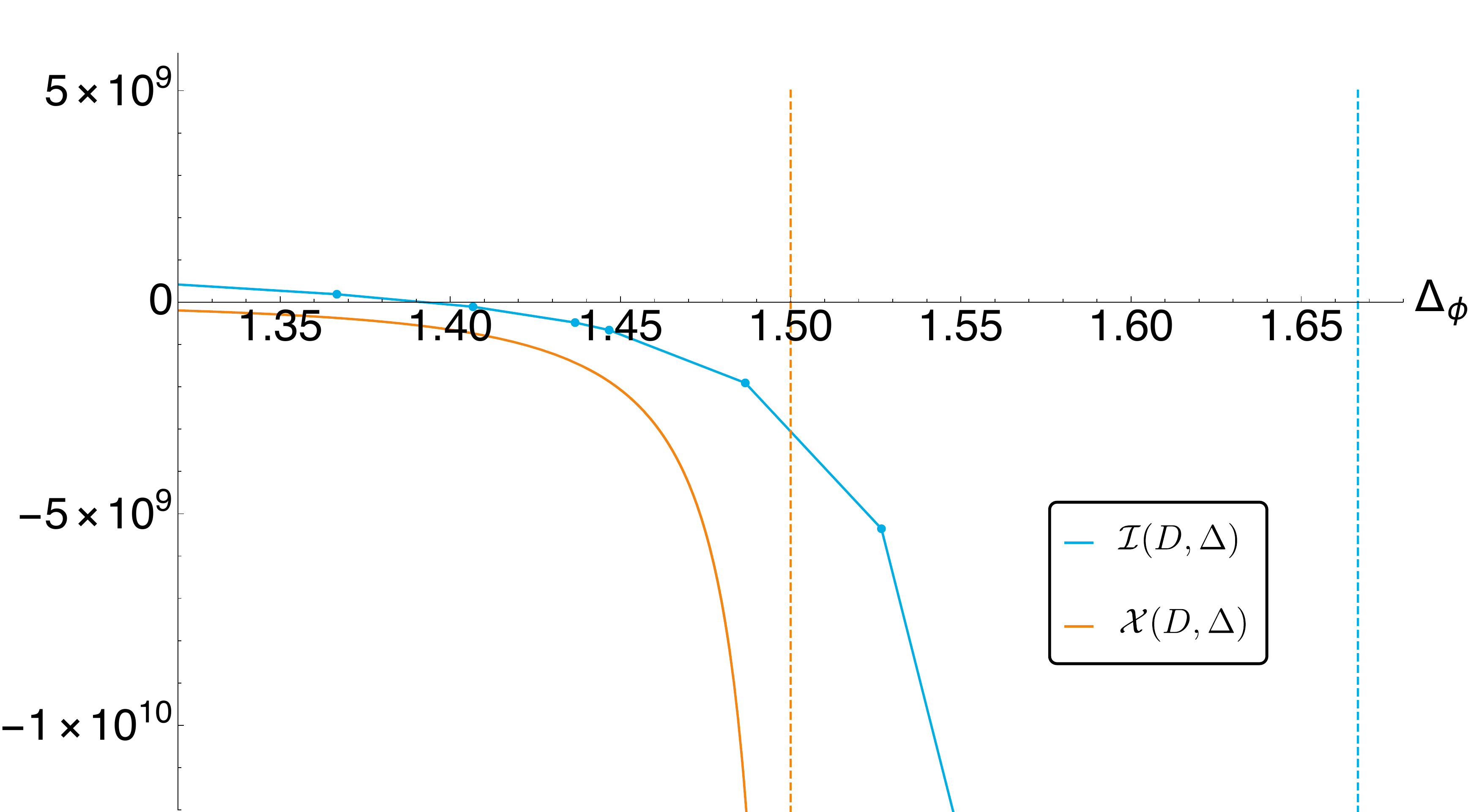}
\includegraphics[width=10cm]{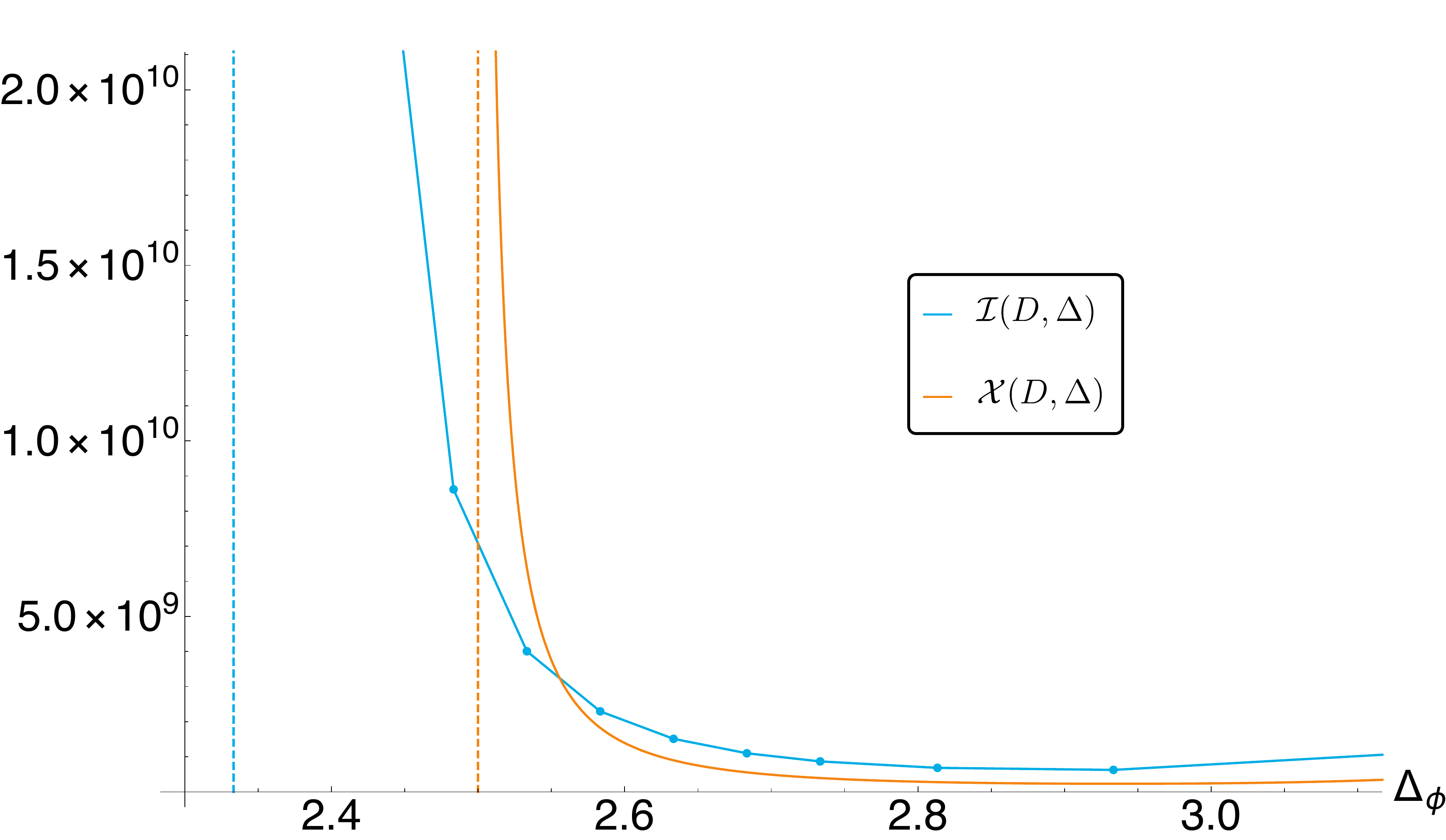}
\end{center} 
\caption{Numerical solutions of the bootstrap equations in $D=5$ and $D=7$ respectively.
The dashed lines indicate poles of $\mathcal{I}$ and $\mathcal{X}$ respectively. }\label{phi3plot}
\end{figure}
We first try to solve the bootstrap equations at $D=5$. It is well known that $\lambda \phi^3$ describes the Lee-Yang edge singularity at $D<6$ \cite{Fisher:1978pf}. As can be seen in Figure \ref{phi3plot}, $\mathcal{I}(D,\Delta)$ and $\mathcal{X}(D,\Delta)$ never intersect, so that we found no solution that corresponds to this CFT.
This is however not surprising, since the propagator bootstrap equation is based on unitarity conditions \cite{polyakov1969microscopic}, while the Lee-Yang edge singularity is a non-unitary CFT.
To maintain unitarity, we can instead solve the equations at $D=7$ and get the second plot in Figure \ref{phi3plot}.
The intersection of $\mathcal{I}(D,\Delta)$ and $\mathcal{X}(D,\Delta)$ is at $\Delta_{}=2.558$. 
Previously, this quantity was calculated using Gliozzi’s fusion rule truncation bootstrap method \cite{Gliozzi:2013ysa} and one gets $\Delta_{}=2.530$ \cite{Rong:2020gbi, Nakayama2015}. We should emphasise that since the $ g \phi^3$ potential is unstable, $\Delta_{}$ receives instanton contributions and will be a complex number. 
The $g \phi^3$ theory is therefore not the perfect model to apply the bootstrap method. 
Our numerical calculation is performed on a laptop. In general, it takes a few minutes to evaluate one data point.

\section{The $O(N)$ vector models}
\label{ONmodels}
In the previous section, we reviewed the old MP bootstrap and developed an efficient method for solving the bootstrap equations in the $g\phi^3$ theory at leading order.
Let us now move on to the more physically interesting case of $O(N)$ vector models in three dimensions.

In the $O(N)$ vector models, the two-point functions $\braket{\phi(x_1)\phi(x_2)}$, $\braket{F(x_1)F(x_2)}$ and the 1PI vertex $\braket{\tilde{\phi}(x_1) \tilde{\phi}(x_2) \tilde{F}(x_3)}$ are fixed by conformal symmetry to be
\begin{align}
\label{2pt-phi}
\Braket{\phi(x_1)\phi(x_2)} \,&=\, {1 \over (x_{12}^2)^{\Delta_\phi}}\,,
\\[0.35 em]
\label{2pt-F}
\Braket{F(x_1) F(x_2)} \,&=\, {1 \over (x_{12}^2)^{\Delta_F}}\,,
\\[0.35 em]
\label{3pt-phi2-F}
\braket{\tilde{\phi}(x_1) \tilde{\phi}(x_2) \tilde{F}(x_3)} \,&=\, 
{g \over (x_{12}^2)^{(D+\Delta_F)/2 - \Delta_\phi} (x_{23}^2)^{(D-\Delta_F)/2} (x_{31}^2)^{(D-\Delta_F)/2}}\,.
\end{align}
The three point functions $\braket{T\phi\phi}$ and $\braket{TFF}$ are also fixed by conformal symmetry. They take the form of \eqref{T-phi-phi}, with $\Delta$ replaced by $\Delta_{\phi}$ and $\Delta_F$ respectively.

There are three bootstrap equations corresponding to the 1PI vertex $\braket{\tilde{\phi}(x_1) \tilde{\phi}(x_2) \tilde{F}(x_3)}$ and two three-point functions $\braket{T_{\mu\nu}(x_1) \phi(x_2) \phi(x_3)}$, $\braket{T_{\mu\nu}(x_1) F(x_2) F(x_3)}$ respectively (c.f.\cite{Grensing:1978cx} for a detailed derivation of the bootstrap equations in this theory).
They are given by the following skeleton expansions,
\begin{align}\label{Onbooteqns}
\begin{aligned}
\text{
\begin{tikzpicture}[scale=1]
\draw[dashed, color=desycyan, line width=0.8pt] (0,0) -- (90:0.5);
\draw[-, color=desycyan, line width=0.8pt] (0,0) -- (220:0.5);
\draw[-, color=desycyan, line width=0.8pt] (0,0) -- (320:0.5);
\fill[color=desyorange] (0,0) circle (3.0pt);
\end{tikzpicture}
}
\end{aligned}
~~{\color{desycyan}=}&~
\begin{aligned}
\text{
\begin{tikzpicture}[scale=0.9]
\draw[-, color=desycyan, line width=0.8pt] (-30:0.8) -- (90:0.8) -- (210:0.8);
\draw[dashed, color=desycyan, line width=0.8pt] (210:0.8) -- (-30:0.8);
\draw[dashed, color=desycyan, line width=0.8pt] (90:0.8) -- (90:1.1);
\draw[-, color=desycyan, line width=0.8pt] (-30:0.8) -- (320:1.1);
\draw[-, color=desycyan, line width=0.8pt] (210:0.8) -- (220:1.1);
\fill[color=desyorange] (-30:0.8) circle (3.0pt);
\fill[color=desyorange] (90:0.8) circle (3.0pt);
\fill[color=desyorange] (210:0.8) circle (3.0pt);
\end{tikzpicture}
}
\end{aligned}
~{\color{desycyan}+}~
\begin{aligned}
\text{
\begin{tikzpicture}[scale=1]
\draw[-, color=desycyan, line width=0.8pt] (90:1.2) -- (35:0.8) -- (145:0.8) -- (90:1.2);
\draw[-, color=desycyan, line width=0.8pt] (-35:0.8) -- (-145:0.8);
\draw[dashed, color=desycyan, line width=0.8pt] (35:0.8) -- (-35:0.8);
\draw[dashed, color=desycyan, line width=0.8pt] (145:0.8) -- (-145:0.8);
\draw[dashed, color=desycyan, line width=0.8pt] (90:1.2) -- (90:1.5);
\draw[-, color=desycyan, line width=0.8pt] (-145:0.8) -- (215:1.1);
\draw[-, color=desycyan, line width=0.8pt] (-35:0.8) -- (325:1.1);
\fill[color=desyorange] (90:1.2) circle (3.0pt);
\fill[color=desyorange] (35:0.8) circle (3.0pt);
\fill[color=desyorange] (-35:0.8) circle (3.0pt);
\fill[color=desyorange] (145:0.8) circle (3.0pt);
\fill[color=desyorange] (-145:0.8) circle (3.0pt);
\end{tikzpicture}
}
\end{aligned}
~{\color{desycyan}+}~
\begin{aligned}
\text{
\begin{tikzpicture}[scale=1]
\draw[-, color=desycyan, line width=0.8pt]  (35:0.8) -- (90:1.2) -- (145:0.8);
\draw[dashed, color=desycyan, line width=0.8pt] (35:0.8) -- (-35:0.8);
\draw[dashed, color=desycyan, line width=0.8pt] (145:0.8) -- (-145:0.8);
\draw[-, color=desycyan, line width=0.8pt] (145:0.8) -- (-35:0.8);
\draw[-, color=desycyan, line width=0.8pt] (-145:0.8) -- (-145:0.1);
\draw[-, color=desycyan, line width=0.8pt] (35:0.8) -- (35:0.1);
\draw[dashed, color=desycyan, line width=1.0pt] (90:1.2) -- (90:1.5);
\draw[-, color=desycyan, line width=0.8pt] (-145:0.8) -- (215:1.1);
\draw[-, color=desycyan, line width=0.8pt] (-35:0.8) -- (325:1.1);
\fill[color=desyorange] (90:1.2) circle (3.0pt);
\fill[color=desyorange] (35:0.8) circle (3.0pt);
\fill[color=desyorange] (-35:0.8) circle (3.0pt);
\fill[color=desyorange] (145:0.8) circle (3.0pt);
\fill[color=desyorange] (-145:0.8) circle (3.0pt);
\end{tikzpicture}
}
\end{aligned}
~~{\color{desycyan}+\, \cdots}
\\[0.6em]
%%%%%%%%%%%%%%%%
%%%%%%%%%%%%%%%%
\label{boot-phi}
\begin{aligned}
\text{
\begin{tikzpicture}[scale=1]
\draw[-, color=desycyan, line width=0.8pt] (-0.6,0) -- (0.6,0);
\fill[color=white] (0,-0.1) circle (0pt);
\draw[-, color=internationalorange, line width=1.0pt] (-60:0.1) -- (120:0.1) -- (0,0) -- (240:0.1) -- (60:0.1);
\end{tikzpicture}
}
\end{aligned}
~{\color{desycyan}=}&~
\begin{aligned}
\text{
\begin{tikzpicture}[scale=1]
\draw[-, color=desycyan, line width=0.8pt] (0.6,0) arc (0:180:0.6 and 0.52);
\draw[dashed, color=desycyan, line width=0.8pt] (0.6,0) arc (0:-180:0.6 and 0.42);
\draw[-, color=desycyan, line width=0.8pt]  (-0.6,0) -- (-1.1,0);
\draw[-, color=desycyan, line width=0.8pt]  (0.6,0) -- (1.1,0);
\fill[color=desyorange] (-0.6,0) circle (3.0pt);
\fill[color=desyorange] (0.6,0) circle (3.0pt);
\draw[yshift=5.1mm, -, color=internationalorange, line width=1.0pt] (-60:0.1) -- (120:0.1) -- (0,0) -- (240:0.1) -- (60:0.1);
\end{tikzpicture}
}
\end{aligned}
~{\color{desycyan}+}~
\begin{aligned}
\text{
\begin{tikzpicture}[scale=1]
\draw[-, color=desycyan, line width=0.8pt] (0.6,0) arc (0:180:0.6 and 0.52);
\draw[dashed, color=desycyan, line width=0.8pt] (0.6,0) arc (0:-180:0.6 and 0.42);
\draw[-, color=desycyan, line width=0.8pt]  (-0.6,0) -- (-1.1,0);
\draw[-, color=desycyan, line width=0.8pt]  (0.6,0) -- (1.1,0);
\fill[color=desyorange] (-0.6,0) circle (3.0pt);
\fill[color=desyorange] (0.6,0) circle (3.0pt);
\draw[yshift=-4.2mm, -, color=internationalorange, line width=1.0pt] (-60:0.1) -- (120:0.1) -- (0,0) -- (240:0.1) -- (60:0.1);
\end{tikzpicture}
}
\end{aligned}
~{\color{desycyan}+}~
\begin{aligned}
\text{
\begin{tikzpicture}[scale=1]
\draw[-, color=desycyan, line width=0.8pt] (1.2,0) arc (0:180:0.8 and 0.55);
\draw[-, color=desycyan, line width=0.8pt] (0.4,0) arc (0:-180:0.8 and 0.55);
\draw[-, color=desycyan, line width=0.8pt] (-0.4,0) -- (0.4,0);
\draw[dashed, color=desycyan, line width=0.8pt] (-1.2,0) -- (-0.4,0);
\draw[dashed, color=desycyan, line width=0.8pt] (1.2,0) -- (0.4,0);
\draw[-, color=desycyan, line width=0.8pt]  (-1.2,0) -- (-1.6,0);
\draw[-, color=desycyan, line width=0.8pt]  (1.2,0) -- (1.6,0);
\fill[color=desyorange] (-0.4,0) circle (3.0pt);
\fill[color=desyorange] (0.4,0) circle (3.0pt);
\fill[color=desyorange] (-1.2,0) circle (3.0pt);
\fill[color=desyorange] (1.2,0) circle (3.0pt);
\draw[yshift=0mm, -, color=internationalorange, line width=1.0pt] (-60:0.1) -- (120:0.1) -- (0,0) -- (240:0.1) -- (60:0.1);
\end{tikzpicture}
}
\end{aligned}
~{\color{desycyan}+\, \cdots}
\\[1.0em]
%%%%%%%%%%%%%%%%
%%%%%%%%%%%%%%%%
\label{boot-F}
\begin{aligned}
\text{
\begin{tikzpicture}[scale=1]
\draw[dashed, color=desycyan, line width=0.8pt] (-0.6,0) -- (0.6,0);
\fill[color=white] (0,-0.1) circle (0pt);
\draw[-, color=internationalorange, line width=1.0pt] (-60:0.1) -- (120:0.1) -- (0,0) -- (240:0.1) -- (60:0.1);
\end{tikzpicture}
}
\end{aligned}
~{\color{desycyan}=}&~
\begin{aligned}
\text{
\begin{tikzpicture}[scale=1]
\draw[-, color=desycyan, line width=0.8pt] (0.6,0) arc (0:180:0.6 and 0.52);
\draw[-, color=desycyan, line width=0.8pt] (0.6,0) arc (0:-180:0.6 and 0.42);
\draw[dashed, color=desycyan, line width=0.8pt]  (-0.6,0) -- (-1.2,0);
\draw[dashed, color=desycyan, line width=0.8pt]  (0.6,0) -- (1.2,0);
\fill[color=desyorange] (-0.6,0) circle (3.0pt);
\fill[color=desyorange] (0.6,0) circle (3.0pt);
\draw[yshift=5.2mm, -, color=internationalorange, line width=1.0pt] (-60:0.1) -- (120:0.1) -- (0,0) -- (240:0.1) -- (60:0.1);
\end{tikzpicture}
}
\end{aligned}
~{\color{desycyan}+~{1 \over 2}}
\begin{aligned}
\text{
\begin{tikzpicture}[scale=1]
\draw[-, color=desycyan, line width=0.8pt] (1.2,0) arc (0:180:0.8 and 0.55);
\draw[-, color=desycyan, line width=0.8pt] (0.4,0) arc (0:-180:0.8 and 0.55);
\draw[dashed, color=desycyan, line width=0.8pt] (-0.4,0) -- (0.4,0);
\draw[-, color=desycyan, line width=0.8pt] (-1.2,0) -- (-0.4,0);
\draw[-, color=desycyan, line width=0.8pt] (1.2,0) -- (0.4,0);
\draw[dashed, color=desycyan, line width=0.8pt]  (-1.2,0) -- (-1.82, 0);
\draw[dashed, color=desycyan, line width=0.8pt]  (1.2,0) -- (1.82, 0);
\fill[color=desyorange] (-0.4,0) circle (3.0pt);
\fill[color=desyorange] (0.4,0) circle (3.0pt);
\fill[color=desyorange] (-1.2,0) circle (3.0pt);
\fill[color=desyorange] (1.2,0) circle (3.0pt);
\draw[-, color=internationalorange, line width=1.0pt] (-60:0.1) -- (120:0.1) -- (0,0) -- (240:0.1) -- (60:0.1);
\end{tikzpicture}
}
\end{aligned}
~{\color{desycyan}+\, \cdots}
\end{align}
In these diagrams, the dashed line and solid line denote the field $\phi$ and $F$ respectively, the orange dot denotes the 1PI vertex and the cross symbol denotes either $\braket{T\phi\phi}$ or $\braket{TFF}$, the internal line denotes the two-point function $\braket{\phi\phi}$ or $\braket{FF}$.

\begin{figure}[ht]
\includegraphics[width=8cm]{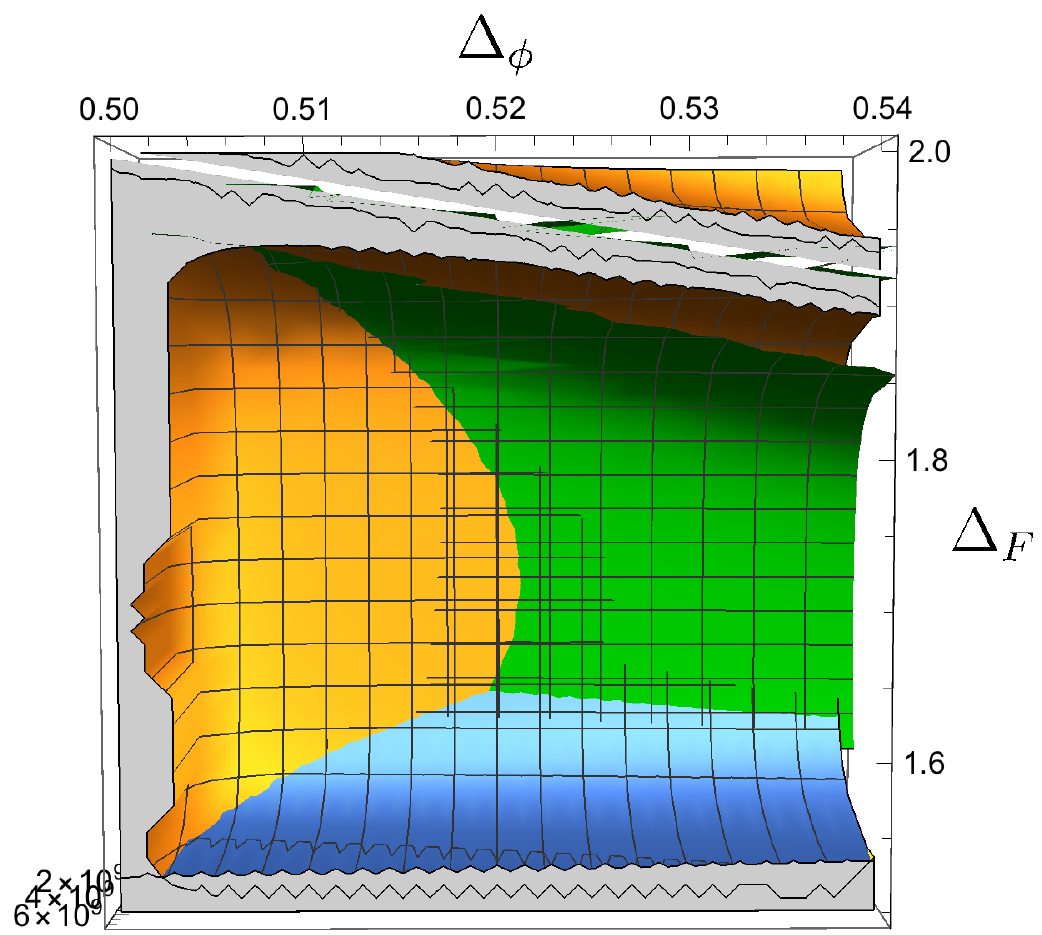}
\includegraphics[width=8cm]{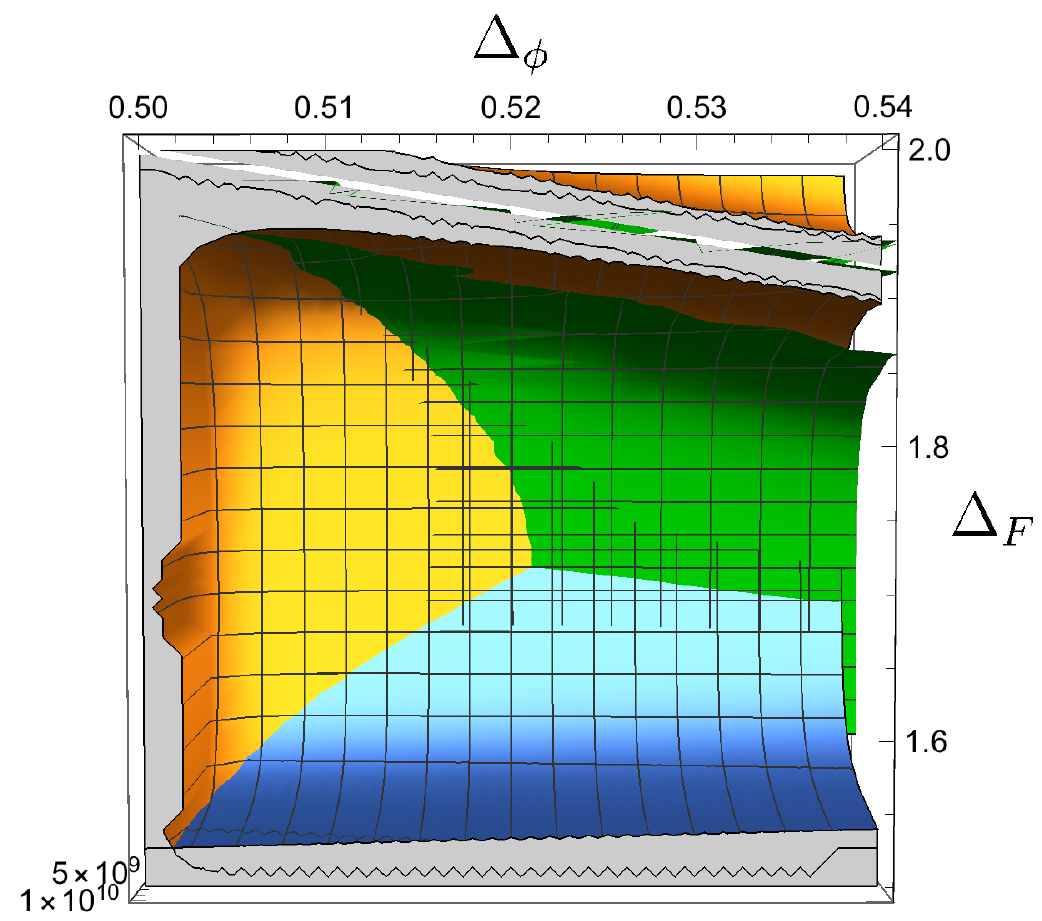}
\caption{Numerical solutions of the bootstrap equations for $O(N)$ vector models. The left plot is for $N=1$ and the right plot is for $N=2$. The green, yellow and blue areas denote $\mathcal{I}(\Delta_{\phi},\Delta_F,D)$, $\mathcal{X}_1(\Delta_{\phi},\Delta_F,D)$ and $\mathcal{X}_2(\Delta_{\phi},\Delta_F,D)$ respectively.}\label{phi4plot}
\end{figure}

It is now straightforward to apply the method described in the previous section to solve the bootstrap equations in the $O(N)$ vector models.
At leading order in the skeleton expansion, the three bootstrap equations in \eqref{Onbooteqns}, \eqref{boot-phi} and \eqref{boot-F} can be written as
\begin{align}\label{On-booteqns-I-facctor}
1 &= g^2\, \mathcal{I}(\Delta_{\phi},\Delta_F,D)+\cdots,\nn\\[0.3 em]
1 &= g^2 \mathcal{X}_1(\Delta_{\phi},\Delta_F,D)+\cdots, \nn\\[0.3 em]
1 &= g^2 N \mathcal{X}_2(\Delta_{\phi},\Delta_F,D)+\cdots.
\end{align}
The extra factor of $N$ in the third equation comes from the $\phi$-loop of the corresponding bootstrap equation. The calculation of  $\mathcal{I}(\Delta_{\phi},\Delta_F,D)$, $\mathcal{X}(\Delta_{\phi},\Delta_F,D)$ and $\mathcal{I}(\Delta_{\phi},\Delta_F,D)$ are analogous what we did for the $g \phi^3$ theory, we skip the details here. The final results are given in \eqref{ONvertexconstant}, \eqref{ONppconstant1} and \eqref{ONppconstant2} respectively. 
We now try to solve the equations in the large $N$ limit, as was done in \cite{Vasiliev:1981yc,vasil19821}.
Setting $D=3$ and $\Delta_{\phi}=\frac{1}{2}+c_1 \frac{1}{N}, \Delta_{F}=2-c_2 \frac{1}{N}$, we get
\begin{align}
\mathcal{I}(\Delta_{\phi},\Delta_F,D) &= -\frac{64 \pi ^6 N^3}{\left(c_1-\frac{c_2}{2}\right){}^3}+\cdots,\nn\\[0.2 em]
\mathcal{X}_1(\Delta_{\phi},\Delta_F,D) &= \frac{64 \pi ^6 N^3}{3 c_1 \left(c_1-\frac{c_2}{2}\right)^2}+\cdots,\nn\\[0.2 em]
N \mathcal{X}_2(\Delta_{\phi},\Delta_F,D) &= \frac{16 \pi ^8 N^3}{\left(c_1-\frac{c_2}{2}\right)^2}+\cdots.
\end{align}
The bootstrap equations have a solution with 
\be
c_1=\frac{4}{3 \pi ^2}\,,\quad  c_2=\frac{32}{3 \pi ^2}\,.
\ee
We can also directly solve them numerically at finite $N$ and get Figure \ref{phi4plot}.
More explicitly, we found
\bea
&& N=1\,,\quad \Delta_{\phi}= 0.52\,,\quad \Delta_{F}=1.62\,,\nn\\
&& N=2\,,\quad \Delta_{\phi}= 0.521\,,\quad \Delta_{F}=1.705\,,\nn\\
&& N=3\,,\quad \Delta_{\phi}= 0.52\,,\quad \Delta_{F}=1.765\,,\nn\\
&& N=5\,,\quad
 \Delta_{\phi}= 0.5168\,,\quad \Delta_{F}=1.83\,,\\
&& N=8,\quad
 \Delta_{\phi}= 0.5126\,,\quad \Delta_{F}=1.882\,.
\eea
This can be compared with 
\bea
&& N=1\,,\quad \Delta_{\phi}=  0.5181489(10)\,,\quad \Delta_{F}= 1.412625(10)\,,\nn\\
&& N=2\,,\quad \Delta_{\phi}= 0.519088(22)\,,\quad \Delta_{F}=1.51136(22)\,,\nn\\
&& N=3\,,\quad \Delta_{\phi}= 0.518936(67)\,,\quad \Delta_{F}=1.59488(81)\,,\nn\\
&& N=5\,,\quad
 \Delta_{\phi}=0.516907\,,\quad \Delta_{F}=1.72788\,,\\
&& N=8\,,\quad
 \Delta_{\phi}= 0.513312\,,\quad \Delta_{F}=1.83188\,.
\eea
The $N=1, 2, 3$ data comes from the numerical bootstrap results in \cite{Kos:2016ysd,Chester:2019ifh,Chester:2020iyt}. The $N=5, 8 $ data comes from the two-sided Pad\'e re-summation of the large $N$ results \cite{Vasiliev:1981yc,vasil19821} together with the $N=1, 2, 3$ data, which gives
\begin{align}
\Delta_{\phi} &= \frac{\frac{129.172}{N^3}+\frac{27.7292}{N^2}+\frac{6.3862}{N}+1}{\frac{250.292}{N^3}+\frac{52.5428}{N^2}+\frac{12.232}{N}+2}\,,
\nn\\[0.3 em]
\Delta_{F} &= \frac{-\frac{1.75813}{N^3}+\frac{14.4595}{N^2}-\frac{2.87073}{N}+2}{\frac{8.26992}{N^2}-\frac{0.894986}{N}+1}\,.
\end{align}

In general, the solutions for $\Delta_{\phi}$ seem reasonable and fair well when compared to other approaches.  The solutions for $\Delta_{F}$ on the other hand contain larger errors, however, we remind the reader that we are working at leading order in the skeleton expansion. It is likely that including higher-order contributions will improve the result.

\section{Conclusions and discussion}\label{Discussionsection}

We have revisited the old Migdal-Polyakov bootstrap equations in light of modern techniques for evaluating multi-loop integrals. Our main result is a framework for computing the generalized Feynman integrals that appear in the MP formalism.
We demonstrated the use of our method by solving the bootstrap equations in $\phi^3$ and $\phi^4$ theories at LO in the skeleton expansion.
At this order, while the propagator equations can be solved analytically, non-trivial three-loop generalized Feynman integrals are involved in the vertex equations.
To solve these integrals, we first derived their MB representations and analytically continued the resulting MB integrals to the region of physical interest.
We then evaluated them numerically using the numerical integration program \texttt{Cuba}.
We found non-trivial solutions corresponding to the $\phi^3$ theory in seven dimensions and the $O(N)$ vector models in three dimensions. One of the merits of these results is that we work directly for integer values of the spacetime dimensions.

It would be highly desirable to push our approach to higher orders.
At NLO, while only a small number of diagrams are involved, their computation is significantly more challenging.
One of the reasons is that the number of integration loops (and propagators) grows quickly.
For example, consider the following diagram for the vertex bootstrap equation in $\phi^3$:
\begin{align*}
\text{
\begin{tikzpicture}[scale=1]
\draw[-, color=desycyan, line width=0.8pt]  (35:0.8) -- (90:1.2) -- (145:0.8);
\draw[-, color=desycyan, line width=0.8pt] (35:0.8) -- (-35:0.8);
\draw[-, color=desycyan, line width=0.8pt] (145:0.8) -- (-145:0.8);
\draw[-, color=desycyan, line width=0.8pt] (145:0.8) -- (-35:0.8);
\draw[-, color=desycyan, line width=0.8pt] (-145:0.8) -- (-145:0.1);
\draw[-, color=desycyan, line width=0.8pt] (35:0.8) -- (35:0.1);
\draw[-, color=desycyan, line width=1.0pt] (90:1.2) -- (90:1.5);
\draw[-, color=desycyan, line width=0.8pt] (-145:0.8) -- (215:1.2);
\draw[-, color=desycyan, line width=0.8pt] (-35:0.8) -- (325:1.2);
\fill[color=desyorange] (90:1.2) circle (3.0pt);
\fill[color=desyorange] (35:0.8) circle (3.0pt);
\fill[color=desyorange] (-35:0.8) circle (3.0pt);
\fill[color=desyorange] (145:0.8) circle (3.0pt);
\fill[color=desyorange] (-145:0.8) circle (3.0pt);
\end{tikzpicture}
}
\end{align*}
At face value this is a 12-loop integral, however six internal coordinates are easily integrated using the star-triangle formula. The resulting 6-loop generalized Feynman integral can be analyzed using our methods.
We derived its MB representation and found that the resulting MB integral has 16 integration variables.
Alternatively, we also derived the corresponding Feynman/Lee-Pomeransky parametric representation that is a 13-fold integral (with a delta function constraint). This is still a challenging task.
A standard method, the sector decomposition technique \cite{Hepp:1966eg,Speer:1975dc,Binoth:2000ps,Binoth:2003ak}, can be used to analytically continue the integral to the physical region where the integral is not convergent.
However, a direct use of the sector decomposition method at NLO looks clumsy and leads to massive computations.
A better understanding of generalized Feynman integrals is still needed, for example, exploiting symmetries of our integrals may simplify our computation. We hope to come back to this in the future.

It is well known that perturbative techniques such as the $\epsilon$-expansion give asymptotic series for physical quantities such as $\Delta_{\phi}$ and $\Delta_{F}$. The series are divergent and proper re-summation is necessary to obtain meaningful results.
This is because of the existence of a branch cut for physical quantities at $\lambda<0$.
In the case of QED and $\lambda \phi^4$ theory, this is also related to the instability of the theory at negative coupling\cite{Dyson:1952tj,simon1970coupling,loeffel1969pade} (see also the review \cite{dunne2002perturbative}).
Such a divergence was proven by explicit calculation in \cite{hurst1952example,hurst1952enumeration,thirring1953divergence,petermann1953divergence} for $\phi^3$ theory.
The divergence is partially due to the rapid growth of the number of Feynman diagrams at higher loops.
Since the MP bootstrap equations are based on a skeleton expansion, the number of generalized Feynman diagrams are smaller.
We might therefore hope that solving the bootstrap equations by successively including higher loops gives us a convergent series.
This brings up the interesting question:~\textit{do the bootstrap equations from the skeleton expansion give a convergent answer which correctly describes the data of the conformal field theory?} Our work can be viewed as a first step towards answering this question, however we emphasize that without including the contribution from higher loop diagrams we cannot make any precise statements.
In any case, it is encouraging to get solutions that give reasonable results.
Hopefully development of efficient techniques to deal with generalized Feynman integrals continues, and the MP bootstrap comes back as an active method to study conformal field theories.

\section*{Acknowledgments}
We would like to thank Wenliang Li, Georgios Papathanasiou, Volker Schomerus, Ning Su and Xiaoran Zhao for helpful discussions. J.\,R.~would also like to thank Valerii L.~Pokrovsky for the lectures on ``Phase transition and renormalization group flow'' given at Texas A\&M University during the fall semester of 2013.
The work of P.\,L.~and J.\,R.~was supported by the Deutsche Forschungsgemeinschaft (DFG) through the Emmy Noether research group ``The Conformal Bootstrap Program'' project number 400570283.
This work was also supported by the DFG under Germany’s Excellence Strategy (EXC 2121) `Quantum Universe’ (No.\,390833306).

\appendix
\section{A brief review of Feynman integrals}\label{sec-app-FI}

In this appendix, we provide a brief introduction to (generalized) Feynman integrals, which is a fundamental tool throughout our work.

We consider the following generalized Feynman integral defined in $D$-dimensional Euclidean space
\begin{align}\label{app-feyn-int-example}
I \,=\, \Bigg(\int \prod_{i=m+1}^{L+m} {d^D\! x_i \over \pi^{D/2}}\Bigg)  {1 \over D_1^{\nu_1} \cdots  D_n^{\nu_n}},
\end{align}
where inverse propagators have the form $D_i \equiv (x_a {-} x_b)^2$, $\nu_i \in \mathbb{R}^+$.
Here $x_i$ with $m < i \leqslant L+m$ denote $L$ integration (loop) coordinates while $x_i$ with $1 \leqslant i \leqslant m$ are the external coordinates.

In general, it is convenient to evaluate Feynman integrals with parametric representation.
Let us perform the following parameterization for each propagator
\begin{align}\label{}
{1 \over A^{a}} \,=\, {1 \over \Gamma(a)} \int_0^\infty d\alpha\, \alpha^{a - 1}\,e^{-\alpha A},
\end{align}
with $\operatorname{Re}(A)>0$ and $\operatorname{Re}(a)>0$.
Then the integral \eqref{app-feyn-int-example} becomes
\begin{align}\label{app-integral-scalar}
I \,=\, \left(\int \prod_{i=m+1}^{L+m} {d^D\! x_i\over \pi^{D/2}} \right)
\left(\prod_{j=1}^n {1 \over \Gamma(\nu_j)} \int_0^\infty d\alpha_j\, \alpha_j^{\nu_j - 1}\right)
\exp\!\Big[\! - A_{ij}\, x_i \cdot x_j + 2B_i\cdot x_i + Q\Big]
\end{align}
where $A$ is a $L\times L$ matrix with scalar products of external coordinates as entries and $B$ is a $L$-vector with the linear combinations of vectors $x_j^\mu$ ($1 \leqslant j \leqslant m$) as entries in the exponential.
Performing Gaussian integrals for all integration variables $x_i$, one obtains
\begin{align}\label{Schwinger-rep} 
I \,&=\, {1 \over \prod_{i=1}^{n} \Gamma(\nu_i)}
\Bigg(\prod_{i=1}^{n} \int_0^\infty d\alpha_i\, \alpha_i^{\nu_i - 1} \Bigg)
{\exp\left(-{\mathcal{F} \over \mathcal{U}}\right) \over \mathcal{U}^{D/2}  },
\end{align}
where $\mathcal{U}$ and $\mathcal{F}$ are the first and second Symanzik polynomials \cite{Bogner:2010kv}.
They both are homogenous polynomials in Feynman parameters and are completely determined by the topology of integral.
Algebraically, they can be evaluated via
\begin{align}\label{} 
\mathcal{U} \,=\, \det A, \quad \mathcal{F} \,=\, \big(\det A\big) \big(Q + B A^{-1}B\big).
\end{align}
From the Schwinger representation in \eqref{Schwinger-rep}, by inserting the identity $1=\int d\eta\, \delta(\eta - \alpha)$ with $\alpha=\sum_i\alpha_i$, rescaling all parameters according to $\alpha_i\to\eta\alpha_i$ and finally integrating over $\eta$, one immediately obtains
\begin{align}\label{Feymannpara} 
I \,&=\, {\Gamma(\nu - LD/2) \over \prod_{i=1}^{n} \Gamma(\nu_i)}
\Bigg(\prod_{i=1}^{n} \int_0^\infty d\alpha_i\, \alpha_i^{\nu_i - 1} \Bigg)
\delta\big(1 - \alpha\big)
{\mathcal{U}^{\nu - (L+1)D/2} \over  \mathcal{F}^{\nu - LD/2}},
\end{align}
where $\nu=\sum_{i=1}^{n} \nu_i$.
It is called the Feynman representation.
Interestingly, Lee and Pomeransky found the following alternative form \cite{Lee:2013hzt}
\begin{align}\label{Lee-para} 
I \,&=\, {\Gamma(D/2) \over \Gamma((L+1)D/2 - \nu)\prod_{i=1}^{n} \Gamma(\nu_i)}
\Bigg(\prod_{i=1}^{n} \int_0^\infty d\alpha_i\, \alpha_i^{\nu_i - 1} \Bigg)
\big(\mathcal{F} + \mathcal{U}\big)^{-D/2}.
\end{align}
Using the same procedure of deriving \eqref{Feymannpara} from \eqref{Schwinger-rep}, one can easily show the equivalence between \eqref{Lee-para} and \eqref{Feymannpara} or \eqref{Schwinger-rep}.

All integrals appear in MP bootstrap equations belong to the following special type of generalized Feynman integrals
\begin{align}\label{n-star-1loop}
I_n \,&=\, \int d^Dx\, {1 \over \prod_{i=1}^{n} [(x {-} x_i)^2]^{\delta_i} }
\quad\text{with}\quad \sum_i \delta_i = D,
\end{align}
where $\{x_1,x_2,x_3\}$ are the external coordinates.
In the case of $n =3$, one can exactly evaluate the integrals and get the so-called {\it star-triangle formula}, i.e.\,\eqref{Gstartriangle}.
This formula was originally derived by Symanzik in \cite{Symanzik:1972wj}.
He also generalized the star-triangle formula to the $n$-star integral \eqref{n-star-1loop} in the case of $n>3$.

Using the Schwinger parameterization, one can write our $n$-star integral \eqref{n-star-1loop} as
\begin{align}\label{symanzik-eq2}
I_n \,&=\, {\pi^{D/2} \over \prod_i \Gamma(\delta_i)}\,
\Bigg(\prod_{i=1}^{n} \int_0^\infty d\alpha_i\, \alpha_i^{\delta_i - 1}\Bigg)
{1 \over \big(\sum_i \lambda_i \alpha_i\big)^{D/2}}\, 
\exp\!\bigg[ \! - {1 \over \sum_i \lambda_i \alpha_i} \sum_{i<j} \alpha_i\alpha_j x_{ij}^2 \bigg],
\end{align}
where $\lambda_i\ge 0$ for any $i$ and $\sum_i \lambda_i^2>0$.
A key observation is that this formula is independent of $\lambda_i\ge 0$, $\forall i$.
To proceed, we note that the exponential function $e^{-z}$ can be represented in terms of Gamma function via the {\it inverse Mellin transformation} formula
\begin{align}\label{inverse-Mellin-minus}
e^{-x} 
\,=\, {1 \over 2\pi i}\int_{c - i\infty}^{c + i\infty} ds\, x^{s}\, \Gamma(-s),
\end{align}
where $c<0$ and the complex function $x^{s}$ are on the principal branch, i.e.~$|\arg x|<\pi/2$.
Finally, the $n$-star Feynman integral \eqref{n-star-1loop} can be written in a {\it Mellin-Barnes} form,
\begin{align}\label{symanzik-mb}
\int {d^Dx \over \prod_{i=1}^{n} [(x {-} x_i)^2]^{\delta_i} }
\,=\, {\pi^{D/2} \over \prod_i \Gamma(\delta_i)}\,
\Bigg(\prod_{\alpha = 1}^{n(n-3)/2}\int_{-i\infty}^{i\infty} ds_\alpha\Bigg)
\Bigg(\raisemath{0.3 em}{\prod_{1\le i < j \le n} } {\Gamma(\Delta_{ij} + \mathcal{S}_{ij}) \over (x_{ij}^2)^{\Delta_{ij} + \mathcal{S}_{ij}}} \Bigg),
\end{align}
with
\begin{align}
\mathcal{S}_{ij} \,\equiv\, \sum_\alpha c_{ij,\alpha} s_\alpha,
\end{align}
where $\Delta_{ij}$ satisfy
\begin{align}\label{}
\sum_{j\ne i} \Delta_{ij} \,=\, \delta_i,
\end{align}
and the real $c_{ij,a}$ obey
\begin{align}\label{}
|\det c_{ij,\alpha}| \,=\, 1, \qquad
\sum_{j\ne i} c_{ij,a} \,=\, 0 
\quad\text{with}\quad
c_{ii,a} \,=\, 0,~~ c_{ij,a} \,=\, c_{ji,a}.
\end{align}
It is worth emphasizing that the integration paths in \eqref{symanzik-mb} have to be along the imaginary axis with real parts such that the real parts of the arguments
of all $\Gamma$ functions are positive.

One of choices for $\Delta_{ij}$ may take
\begin{align}\label{}
\qquad \Delta_{ij} \,=\,  {\delta_i+\delta_j \over n-2} - \frac{\sum_a \delta_{a}}{(n{-}1)(n{-}2)}.
\end{align}
When $n=3$, one can easily check that the formula \eqref{symanzik-mb} is immediately reduced to the star-triangle formula \eqref{Gstartriangle}.

In the case of $n=4$, we may take 
\begin{align}\label{}
 \mathcal{S} =\left(
\begin{array}{cccc}
 0 & s_2 & s_1 & -s_1-s_2 \\
 s_2 & 0 & -s_1-s_2 & s_1 \\
 s_1 & -s_1-s_2 & 0 & s_2 \\
 -s_1-s_2 & s_1 & s_2 & 0 \\
\end{array}
\right).
\end{align}

\section{The propagator equations}\label{Stresstensoreqn}
In this appendix, we provide detail of analytically solving the leading-order propagator bootstrap equations.

The stress-energy tensor $T_{\mu\nu}$ is a short representation of the conformal group, its descendant $\partial^{\mu}T_{\mu\nu}$ is a null state.
The representation theory of conformal group tells us that when multiplets shortening happens, the null state $\partial^{\mu}T_{\mu\nu}$ is also a conformal primary. 
Notice the relative coefficients between the two terms of the \eqref{WardID} can be fixed by requiring the right-hand side to transforms like the three point function of a conformal primary vector field and two conformal primary scalars.
The three point funtion of $T_{\mu\nu}$ and two non-identical scalar is allowed by conformal symmetry
\begin{align}
 \langle T_{\mu\nu}(x_1)\phi_A(x_2)\phi_B(x_3) \rangle=f_{ABT}\times \frac{Z_{\mu}Z_{\nu}-\frac{1}{D}\delta_{\mu\nu} Z_{\rho}Z^{\rho}}{(x_{12}^2)^{\frac{D+\Delta_A-\Delta_B-2}{2}}(x_{13}^2)^{\frac{D+\Delta_B-\Delta_A-2}{2}}(x_{23}^2)^{\frac{\Delta_A+\Delta_B-D+2}{2}}}.
\end{align}
It is however not allowed by Ward identity, which implies that the $f_{ABT}$ vanishes for a local conformal field theory.
More explicitly, 
\begin{align}\label{derivativeT}
\langle \partial_{\nu} & T^{\mu\nu}(x_1)\phi_A(x_2)\phi_B(x_3) \rangle
\nonumber\\
&=-\frac{(D-1)(\Delta_A-\Delta_B)f_{ABT}}{D}\times \frac{Z^{\mu}}{(x_{12}^2)^{\frac{D+\Delta_{A}-\Delta_{B}}{2}}(x_{13}^2)^{\frac{D+\Delta_{B}-\Delta_{A}}{2}}(x_{23}^2)^{\frac{\Delta_{A}+\Delta_{B}-D}{2}}},
\end{align}
indeed looks like a three point function involving two scalars and a conformal primary vector field with scaling dimension $D+1$.
Setting $\Delta_{B}=\Delta_{\phi}$, $\Delta_{A}=\Delta_{\phi}+\epsilon$ and taking the limit $\epsilon\to 0$, we get the familiar differential form of Wald identity \eqref{WardID}.

Let us start with the propagator bootstrap equation \eqref{ppequations} in $g\phi^3$ theory.
Below we show that the right-hand side of the equation at LO has the following form
\begin{align}\label{ppboot1}
&\int_{456789}
\langle \phi (6) \phi (7)\rangle  
\langle \phi_A (8) \phi_A (2)\rangle
\langle \phi_B (9) \phi_B (3)\rangle
\langle \tilde{\phi }(4) \tilde{\phi }(6) \tilde{\phi }_A(8)\rangle
\langle \tilde{\phi }(5) \tilde{\phi }(7) \tilde{\phi }_B(9)\rangle
\langle \phi (4) \phi (5) T_{\mu \nu }(1)\rangle
\nonumber\\
&=\mathcal{X}(D,\Delta,\Delta_A,\Delta_B)\, \langle T_{\mu\nu}(1) \phi_A(2) \phi_B(3)\rangle,
\end{align}
where we used the short-handed notations, $\int_{i\cdots j} \equiv \int d^D\! x_i\cdots d^D\! x_j$, $\phi(i) \equiv \phi(x_i)$.
To use the Ward identity, here we are using the prescription as follows:~we consider $\braket{T^{\mu\nu}(x_1)\phi_A(x_2)\phi_B(x_3)}$ in \eqref{ppboot1}, and will set $\Delta_B=\Delta_\phi$ and $\Delta_{A}=\Delta_{\phi}+\epsilon$ and take the limit $\epsilon\rightarrow 0$ in the final step.

Now let us apply the Ward identity for the equation.
For \eqref{ppboot1},
\begin{align}\label{ppboot1-2}
&\int_{456789}
\langle \phi (6) \phi (7)\rangle  
\langle \phi_A (8) \phi_A (2)\rangle
\langle \phi_B (9) \phi_B (3)\rangle
\langle \tilde{\phi }(4) \tilde{\phi }(6) \tilde{\phi }_A(8)\rangle
\langle \tilde{\phi }(5) \tilde{\phi }(7) \tilde{\phi }_B(9)\rangle
\langle \partial^\mu T_{\mu \nu }(1) \phi (4) \phi (5) \rangle
\nonumber\\
&=\mathcal{X}(D,\Delta,\Delta_A,\Delta_B)\, \langle \partial^\mu T_{\mu\nu}(1) \phi_A(2) \phi_B(3)\rangle,
\end{align}
While the second line in \eqref{ppboot1-2} is just \eqref{derivativeT} up to the factor $\mathcal{X}$, the first line of \eqref{ppboot1-2} is given by
\begin{align}\label{ppboot1-3}
\mathcal{C}_{D}
&
\int_{56789} \langle \phi_A (8) \phi_A (2)\rangle  \langle \phi_B (9) \phi_B (3)\rangle
\nonumber\\
&\times
\Big[
\langle \partial_{\nu} \phi (1)\phi (5)\rangle
\langle \tilde{\phi }(1) \tilde{\phi }(6) \tilde{\phi }_A(8)\rangle
\langle \tilde{\phi }(5) \tilde{\phi }(7) \tilde{\phi }_B(9)\rangle
\langle \phi (6) \phi (7)\rangle
\\
&\quad
-\frac{\Delta}{D}\partial_{\nu}\big(
\langle \phi (1) \phi (5)\rangle
\langle \phi (6) \phi (7)\rangle
\langle \tilde{\phi }(1) \tilde{\phi }(6) \tilde{\phi }_A(8)\rangle
\langle \tilde{\phi }(5) \tilde{\phi }(7) \tilde{\phi }_B(9)\rangle
\big)
+\big\{ 8\leftrightarrow 9, A\leftrightarrow B \big\}\Big]
\nonumber
\end{align} 
with $\mathcal{C}_{\Delta} = 2 \sqrt{\pi}\,\Gamma\big(\frac{D-1}{2}\big)/\Gamma\big(\frac{D}{2}\big)$.
Interestingly, all integrals \eqref{ppboot1-3} can be evaluated by analytically.
First, for the second term in the squared bracket in \eqref{ppboot1-2}, a straightforward use of the star-trianglar formula gives
\begin{align}\label{ppboot1-wi-2}
-\frac{\Delta}{D} &\partial_{\nu}
\int_{567}
\langle \phi (1) \phi (5)\rangle
\langle \phi (6) \phi (7)\rangle
\langle \tilde{\phi }(1) \tilde{\phi }(6) \tilde{\phi }_A(8)\rangle
\langle \tilde{\phi }(5) \tilde{\phi }(7) \tilde{\phi }_B(9)\rangle
\nonumber\\
&= -\frac{\Delta}{D}\partial_{\nu} \frac{g^2 \kappa_1 \kappa_2 \kappa_3}{(x_{18}^2)^{D+\Delta_B-\Delta_A \over 2} (x_{19}^2)^{D+\Delta_A-\Delta_B \over 2} (x_{89}^2)^{D-\Delta_A-\Delta_B \over 2}},
\end{align}
with
\bea
&&\kappa_1=\kappa(2 \Delta ,D-\Delta _B,\Delta _B+D-2 \Delta),\nn\\
&&\kappa_2=\kappa(2 \Delta,D-\Delta _A, \Delta _A+D-2 \Delta ),\nn\\
&&\kappa_3=\kappa(\Delta _A+\Delta _B,2 \Delta -\Delta _A,2 D -\Delta _B-2 \Delta).
\eea
Now let us compute the first term in the squared bracket.
Using the star-trianglar formula iteratively yields
\begin{align}\label{}
&\int_{567}
\langle \partial_{\nu} \phi(1)\phi(5)\rangle
\langle \tilde{\phi }(1) \tilde{\phi }(6) \tilde{\phi }_A(8)\rangle
\langle \tilde{\phi }(5) \tilde{\phi }(7) \tilde{\phi }_B(9)\rangle
\langle \phi (6) \phi (7)\rangle
\nonumber\\
&=\int_{567}
\bigg(\partial_{\nu}\frac{1}{x_{15}^{2 \Delta } x_{95}^{D-\Delta _B} x_{57}^{\Delta _B+D-2 \Delta }}\bigg)
\frac{g^2}{x_{67}^{2 \Delta } x_{18}^{D-\Delta _A} x_{26}^{D-\Delta _A} x_{16}^{\Delta _A+D-2 \Delta } x_{97}^{D-\Delta _B}}
\nonumber\\
&=\int_{7}
\bigg(\partial_{\nu}\frac{1}{x_{19}^{2 \Delta -\Delta _B}}\bigg)\frac{g^2\,\kappa_1 \kappa_2}{x_{78}^{2 \Delta -\Delta _A} x_{17}^{\Delta _A+\Delta _B} x_{18}^{-\Delta _A+2 D-2 \Delta } x_{79}^{-\Delta _B+2 D-2 \Delta }}
\nonumber\\
&\quad
+\frac{\Delta _B}{\Delta _A+\Delta _A}\bigg(\partial_{\nu}\frac{1}{x_{78}^{2 \Delta -\Delta _A} x_{17}^{\Delta _A+\Delta _B} x_{79}^{-\Delta _B+2 D-2 \Delta }}\bigg)\frac{g^2\,\kappa_1 \kappa_2}{x_{19}^{2 \Delta -\Delta _B} x_{18}^{-\Delta _A+2 D-2 \Delta }}
\label{step28}
\\
&= \bigg(\partial_{\nu}\frac{1}{x_{19}^{2 \Delta -\Delta _B}}\bigg)\frac{g^2\,\kappa_1 \kappa_2\kappa_3}{x_{18}^{-\Delta _A+2 D-2 \Delta } x_{19}^{\Delta _A+D-2 \Delta } x_{18}^{\Delta _B-D+2 \Delta } x_{89}^{-\Delta _A-\Delta _B+D}}
\nonumber\\
\label{ppboot1-wi-1}
&\quad
+\frac{\Delta _B}{\Delta _A+\Delta _B} \bigg(\partial_{\nu}\frac{1}{x_{19}^{\Delta _A+D-2 \Delta } x_{18}^{\Delta _B-D+2 \Delta }}\bigg)\frac{g^2\,\kappa_1 \kappa_2\kappa_3}{x_{19}^{2 \Delta -\Delta _B} x_{18}^{-\Delta _A+2 D-2 \Delta } x_{89}^{-\Delta _A-\Delta _B+D}},
\end{align}
where $x_{ij}^a$ should be understood as $(x_{ij}^2)^{a/2}$, we have used $f(x)^{a}\, \partial_{x} \big(f(x)^{b}\big) =\frac{b}{a+b}\,\partial_{x} \big(f(x)^{a + b}\big)$ in \eqref{step28}.

By combining the two terms, \eqref{ppboot1-wi-1} and \eqref{ppboot1-wi-2}, we obtain
\begin{align}
&\int_{4567}
\langle  \partial^{\mu}T_{\mu \nu }(1)\phi (4) \phi (5)\rangle
\langle \phi (6) \phi (7)\rangle
\langle \tilde{\phi }(4) \tilde{\phi }(6) \tilde{\phi }_A(8)\rangle
\langle \tilde{\phi }(5) \tilde{\phi }(7) \tilde{\phi }_B(9)\rangle
\nonumber\\
&= \frac{\mathcal{C}_{\Delta}\, g^2\, Z_{\nu}}{x_{18}^{-\Delta _A+\Delta _B+D} x_{19}^{\Delta _A-\Delta _B+D} x_{89}^{-\Delta _A-\Delta _B+D}}\times \bigg(\kappa(\Delta _A+D-2 \Delta ,D-\Delta _A,2 \Delta) 
\nn\\
&\quad\times \kappa(2 \Delta ,D-\Delta _B,\Delta _B+D-2 \Delta) \kappa(\Delta _A+\Delta _B,2 \Delta -\Delta _A,2 D-\Delta _B-2 \Delta) 
\nn\\
&\quad\times \frac{D \Delta  \Delta _A-\Delta  \Delta _A^2+\Delta _B (D-\Delta ) \left(D-\Delta _B\right)}{D \left(\Delta _A+\Delta _B\right)} - \{A\leftrightarrow B\}\bigg).
\end{align}
The right-hand side of the above equations can be written as
\begin{align}\label{simpleppboot}
g^2\,\hat{\mathcal{X}}(D,\Delta,\Delta_A,\Delta_B)\, \langle  T_{\mu \nu }(1)\tilde{\phi}_A (8) \tilde{\phi}_B (9)\rangle
\end{align}
with
\begin{align}\label{chihat}
\hat{\mathcal{X}}(D,\Delta,\Delta_A,
\Delta_B)=\,&\frac{2 \pi^{D/2}}{\Delta_{\phi } \Gamma(\frac{D}{2})}\frac{1}{(\Delta_{A} {-} \Delta_B)}\bigg(\kappa(\Delta _A+D-2 \Delta, D-\Delta_A, 2\Delta) 
\nn\\
&\times \kappa(2\Delta,D-\Delta_B,\Delta_B+D-2 \Delta)\, \kappa(\Delta _A+\Delta _B,2 \Delta -\Delta _A, 2D - \Delta_B - 2\Delta) 
\nn\\
&\times\frac{D \Delta  \Delta_A-\Delta\Delta_A^2 + \Delta_B (D-\Delta ) (D-\Delta _B)}{D(\Delta _A+\Delta _B)} - \{A\leftrightarrow B\}\bigg).
\end{align}
Integrating out further $x_8$ and $x_9$ in \eqref{ppboot1-3} and noting \eqref{derivativeT}, we prove \eqref{ppboot1-3} or \eqref{ppboot1-2}, and obtain explicitly
\begin{align}\label{chinohat}
\mathcal{X}(D,\Delta)=\, &\left(\lim_{\epsilon\rightarrow 0}\hat{\mathcal{X}}(D,\Delta,\Delta,
\Delta+\epsilon)\right)\frac{\left(\Delta _{\phi }-1\right) \Delta _{\phi }}{\left(-\Delta _{\phi }+D-1\right) \left(D-\Delta _{\phi }\right)} \kappa \left(-2 \Delta _{\phi }+2 D-2,2,2 \Delta _{\phi }\right)
\nn\\
&\times \kappa \left(D-2,2 \left(D-\Delta _{\phi }\right)-D+2,2 \Delta _{\phi }\right). 
\end{align}
Taking the $\epsilon\rightarrow0$ limit, we finally obtain
\begin{align}\label{prop-eq-int-res-app}
\mathcal{X}(D,\Delta) =\,& \frac{\pi^{3 D} (\Delta -1)\, \Gamma(\Delta -1)\, \Gamma^3(\frac{\Delta }{2})}{2\Delta D (D-\Delta -1) (D-\Delta ) \Gamma(\frac{D}{2})\, \Gamma^5(\Delta )\, \Gamma(D-\frac{3 \Delta }{2})}
\nn\\
&\times \frac{\Gamma^5(\frac{D}{2}-\Delta)\, \Gamma(\Delta - \frac{D}{2} +1)\, \Gamma(\frac{3 \Delta }{2}-\frac{D}{2})}
{\Gamma(\frac{D}{2}-\Delta +1)\,\Gamma(D-\Delta -1)\, \Gamma^3(\frac{D-\Delta }{2})}
\\
&\times\Big(
D^2\Delta\, \psi\big(\tfrac{\Delta }{2}\big)
- D^2\Delta\, \psi\big(\tfrac{3 \Delta }{2} {-} \tfrac{D}{2}\big)
-2 D^2 - 8 \Delta ^2
- D\Delta^2\, \psi\big(\tfrac{\Delta }{2}\big)
+8\Delta D
\nn\\
&\qquad +D \Delta^2 \,\psi\big(\tfrac{3 \Delta }{2} {-} \tfrac{D}{2}\big)
+\Delta  D (D-\Delta )\, \psi\big(\tfrac{D-\Delta }{2}\big)
+\Delta  D (\Delta -D)\, \psi\big(D {-} \tfrac{3 \Delta }{2}\big)\Big).
\nonumber
\end{align}
$\mathcal{X}(D,\Delta)$ also has a pole like singularity at $\Delta=\frac{D-2}{2}$.

In fact, we find that the same result can be obtained by solving the equation directly (without using the Ward identity).
It is also a good cross-check for our calculation. 
When $\mu\neq\nu$, we have
\begin{align}\label{ppboot}
&\int_{4\cdots 9}  \langle T_{\mu\nu}(x_1)\phi(x_4) \phi(x_7)\rangle \langle\tilde{\phi}(x_2) \tilde{\phi}(x_4)\tilde{\phi}(x_6)\rangle \langle\tilde{\phi}(x_3) \tilde{\phi}(x_5)\tilde{\phi}(x_7)\rangle \langle\phi(x_6) \phi(x_7)\rangle  \nonumber\\
&= -\frac{1}{(D-2)^2}\partial_{\mu}\partial_{\nu} \int_{4\cdots 9}
\frac{1}{
(x_{14}^2 x_{15}^2)^{\frac{D-2}{2}}
(x_{24}^2 x_{26}^2 x_{46}^2 x_{35}^2 x_{57}^2 x_{37}^2)^{\frac{D-\Delta}{2}}
(x_{45}^2)^{\frac{2\Delta-D+2}{2}}
(x_{67}^{2})^{\frac{2\Delta}{2}}}
\nonumber\\
&\quad
+\frac{2 (D{-}1)}{D (D{-}2)^2}\int_{4\cdots 9} \bigg(\partial_{\mu}\partial_{\nu}\frac{1}{(x_{14}^2)^{\frac{D-2}{2}}}\bigg)
\frac{1}{
(x_{15}^2)^{\frac{D-2}{2}}
(x_{45}^2)^{\frac{2\Delta-D+2}{2}}
(x_{67}^{2})^{\frac{2\Delta}{2}}
}
\frac{1}{
(x_{24}^2 x_{26}^2 x_{46}^2 x_{35}^2 x_{57}^2 x_{37}^2)^{\frac{D-\Delta}{2}}
}
\nonumber\\
&\quad
+\int_{4\cdots 9} \bigg(\partial_{\mu}\partial_{\nu} \frac{1}{(x_{15}^2)^{\frac{D-2}{2}}}\bigg)
\frac{1}{
(x_{14}^2)^{\frac{D-2}{2}}
(x_{45}^2)^{\frac{2\Delta-D+2}{2}}
(x_{67}^{2})^{\frac{2\Delta}{2}}
}
\frac{1}{
(x_{24}^2 x_{26}^2 x_{46}^2 x_{35}^2 x_{57}^2 x_{37}^2)^{\frac{D-\Delta}{2}}
}
\end{align}
 We have used the $\alpha=0$ case of the following equation,
\begin{align}\label{sumT}
C\frac{Z_{\mu}Z_{\nu}
-\frac{1}{D}\delta_{\mu\nu} Z_{\rho}Z^{\rho}}{(x_{12}^2)^{\frac{D-2}{2}}(x_{13}^2)^{\frac{D-2}{2}}(x_{23}^2)^{\frac{2\Delta-D+2}{2}}}={-\frac{1}{(D-2)^2}}\times \partial_{\mu}\partial_{\nu} \frac{1}{{(x_{12}^2)^{\frac{D-2}{2}}(x_{13}^2)^{\frac{D-2}{2}}(x_{23}^2)^{\frac{2\Delta-D+2}{2}}}}\nonumber\\+\frac{2 (D-1)}{(D-2)^2 D}\Bigg\{\left( \partial_{\mu}\partial_{\nu} \frac{1}{{(x_{12}^2)^{\frac{D-2-\alpha}{2}}}{(x_{13}^2)^{\frac{\alpha}{2}}}}\right)\frac{1}{(x_{12}^2)^{\frac{\alpha}{2}}(x_{13}^2)^{\frac{D-2-\alpha}{2}}(x_{23}^2)^{\frac{2\Delta-D+2}{2}}}\nonumber\\+\left( \partial_{\mu}\partial_{\nu} \frac{1}{{(x_{12}^2)^{\frac{\alpha}{2}}(x_{13}^2)^{\frac{D-2-\alpha}{2}}}}\right)\frac{1}{(x_{12}^2)^{\frac{D-2-\alpha}{2}}(x_{13}^2)^{\frac{\alpha}{2}}(x_{23}^2)^{\frac{2\Delta-D+2}{2}}}\Bigg\},
\end{align} 
where
\begin{align}\label{Cfunction}
C(D,\alpha)=\frac{-4 \alpha  (\alpha +2)+D^3-4 (\alpha +1) D^2+4 \left(\alpha ^2+3 \alpha +1\right) D}{(D-2)^2 D}.
\end{align}
The formula is valid when $\mu\neq \nu$. $C$ is clearly invariant under $\alpha\rightarrow D-2-\alpha$.
The first term in \eqref{ppboot} can be represented by the following MB representation
\begin{align}\label{}
&M_1=\frac{-1}{(D-2)^2}
\bigg(\partial_{\mu}\partial_{\nu} \frac{\kappa(D-\Delta ,D-\Delta ,2\Delta)}{{(x_{12}^2)^{\frac{D-2}{2}}(x_{13}^2)^{\frac{D-2}{2}}(x_{23}^2)^{\frac{2\Delta-D+2}{2}}}}\bigg)
\int du_4 dv_4 du_5 dv_6\, f(u_4,v_4,u_7,v_7)
\end{align} 
where 
 \begin{align}\label{ffunction}
f(u_4,v_4,u_7,v_7) =\, 
&\frac{
\kappa\big(\tfrac{D+3 \Delta +2 u_4+4 u_7-4}{2} ,\tfrac{D-\Delta -4 u_4-8 u_7-4 v_4-8 v_7+6}{4} ,\tfrac{5 D-5 \Delta +4 v_4+8 v_7+2}{4}\big)
}{\Gamma(\frac{\Delta}{2})\,
\Gamma(\frac{D-2}{2})\
\Gamma(D-\frac{3 \Delta }{2})\,
\Gamma(\frac{D-\Delta }{2})\,
\Gamma(1-\frac{D}{2}+\Delta)\,
\Gamma(\frac{D}{3}+u_4)\,
\Gamma(\frac{D+3 \Delta +12 v_4+6}{12})}
\nonumber\\
&\times
\Gamma\big(\tfrac{D}{3}-\tfrac{\Delta }{2}+u_4\big)
\Gamma\big(\tfrac{\Delta}{2} -\tfrac{D}{6}+u_4\big)
\Gamma\big(\tfrac{\Delta}{4} -\tfrac{D}{12} +u_7-\tfrac{u_4}{2}\big)
\Gamma\big(\tfrac{D-\Delta}{4} + \tfrac{1}{2}u_4 + u_7\big)
\nonumber\\
&\times
\Gamma\big(\tfrac{7D}{12} - \tfrac{3\Delta}{4} - \tfrac{1}{2} + v_4\big) 
\Gamma\big(\tfrac{1}{2} - \tfrac{5D}{12} + \tfrac{3\Delta}{4} + v_4\big) 
\Gamma\big(\tfrac{D-3 \Delta -12 u_4-12 v_4+6}{12}\big)
\nonumber\\ &\times
\Gamma\big(\tfrac{D+3 \Delta +12 u_4-24 u_7+12 v_4-24 v_7+6}{24}\big) 
\Gamma\big(\tfrac{D-\Delta -4 u_4-8 u_7-4 v_4-8 v_7-2}{8}\big)
\nonumber\\ 
&\times
\Gamma\big(\tfrac{D-\Delta +4 v_4+8 v_7+2}{8}\big) 
\Gamma\big(\tfrac{D+3 \Delta -12 v_4+24 v_7-6}{24}\big).
\end{align} 
Similarly, the second term of \eqref{ppboot} is given by
\begin{align}\label{}
M_2=\,& \frac{2 (D-1)}{(D-2)^2 D}
\frac{1}{(x_{23}^2)^{\frac{2\Delta-D+2}{2}}}
\Bigg\{\left( \partial_{\mu}\partial_{\nu} \frac{1}{{(x_{12}^2)^{\frac{D-2-\alpha}{2}}}{(x_{13}^2)^{\frac{\alpha}{2}}}}\right)
\frac{1}{(x_{12}^2)^{\frac{\alpha}{2}}(x_{13}^2)^{\frac{D-2-\alpha}{2}}}
\\
&
+\left( \partial_{\mu}\partial_{\nu} \frac{1}{{(x_{12}^2)^{\frac{\alpha}{2}}(x_{13}^2)^{\frac{D-2-\alpha}{2}}}}\right)\frac{1}{(x_{12}^2)^{\frac{D-2-\alpha}{2}}(x_{13}^2)^{\frac{\alpha}{2}}}\Bigg\}
\int du_4 dv_4 du_5 dv_6\, f(u_4,v_4,u_7,v_7),
\nonumber
\end{align}
with 
\begin{align}\label{}
\alpha=\frac{1}{4} \left(-D+5 \Delta -4 v_4-8 v_7-2\right).
\end{align} 
Now using \eqref{sumT}, we find
\begin{align}
&\Bigg(\prod_{i=4}^{9}\int d^D\! x_i \!\Bigg)  \langle T_{\mu\nu}(x_1)\phi(x_4) \phi(x_7)\rangle \langle\tilde{\phi}(x_2) \tilde{\phi}(x_4)\tilde{\phi}(x_6)\rangle \langle\tilde{\phi}(x_3) \tilde{\phi}(x_5)\tilde{\phi}(x_7)\rangle \langle\phi(x_6) \phi(x_7)\rangle\nonumber\\
&=M_1+M_2
=\mathcal{J}(D,\Delta)\, \langle T_{\mu \nu}(x_1) \tilde{\phi}(x_2) \tilde{\phi}(x_3)\rangle
\end{align}
with 
\begin{align}
\mathcal{J}(D,\Delta) = C(D,\alpha)\,\int du_4 dv_4 du_5 dv_6\, f(u_4,v_4,u_7,v_7),
\end{align}
where $f$ and $C$ are defined in \eqref{ffunction} and \eqref{Cfunction} respectively.
The MB integral in $M_1$ or $M_2$ can be analytically calculated using \texttt{MB.m}.
We found perfect agreement with the result from Ward identity, i.e.\,\eqref{prop-eq-int-res-app}.

The bootstrap equations \eqref{Onbooteqns} in $O(N)$ vector models can be analytically solved at the leading order in a similar manner.
We do not write down the detail and list the results as follows:
\begin{align}\label{ONppconstant1}
\mathcal{X}_1(\Delta_{\phi},\Delta_F,D) =&
\frac{\pi^{3 D+1}\,\Gamma^2\big(\tfrac{\Delta_F}{2}\big)\, \csc\big(\tfrac{D-2 \Delta_{\phi}}{2} \pi\big) \,\Gamma\big(\tfrac{D}{2}-\Delta_F\big)}{\Delta_{\phi}^2\,\Gamma\big(\tfrac{D}{2}\big)\,\Gamma\big(\Delta_F\big) \,\Gamma^2\big(\tfrac{D-\Delta_F}{2}\big) \,\Gamma\big(\tfrac{D}{2} - \Delta_{\phi} + \tfrac{\Delta_F}{2}\big)}
\nn\\
&\times\frac{\Gamma\big(\Delta_{\phi}+1\big) \,\Gamma^3\big(\tfrac{D}{2}-\Delta_{\phi}\big) \,\Gamma\big(\Delta_{\phi}-\tfrac{\Delta_F}{2}\big) \,\Gamma\big(\Delta_{\phi}-\tfrac{D}{2}+\tfrac{\Delta_F}{2}\big)}
{\Gamma^4\big(\Delta_{\phi}\big) \,\Gamma\big(1-\Delta_{\phi}+\tfrac{D}{2}\big) \,\Gamma\big(D-\Delta_{\phi}+1\big) \,\Gamma\big(D-\Delta_{\phi}-\tfrac{\Delta_F}{2}\big)}
\nn\\
&\times\Big(
\big(2 \Delta_{\phi} (D-\Delta_{\phi})-D \Delta_F+\Delta_F^2\big) \psi\big(\tfrac{\Delta_F}{2}\big)
\nn\\
&\qquad +(2 \Delta_{\phi} (D-\Delta_{\phi})-D \Delta_F+\Delta_F^2) \,\psi\big(\tfrac{D-\Delta_F}{2}\big)
\nn\\
&\qquad + \Delta_{\phi} (\Delta_{\phi}-D) \,\psi\big(D -\Delta_{\phi}-\tfrac{\Delta_F}{2}\big)
+\Delta_{\phi} (\Delta_{\phi}-D) \,\psi\big(\Delta_{\phi}-\tfrac{D}{2}+\tfrac{\Delta_F}{2}\big)
\nn\\
&\qquad +(\Delta_{\phi}-\Delta_F) (\Delta_{\phi}-D+\Delta_F) \,\psi\big(\Delta_{\phi}-\tfrac{\Delta_F}{2}\big)
\nn\\
&\qquad +(\Delta_{\phi}-\Delta_F) (\Delta_{\phi}-D+\Delta_F) \,\psi\big(\tfrac{D}{2} - \Delta_{\phi} + \tfrac{\Delta_F}{2}\big)
\nn\\
&\qquad -2 (D-2 \Delta_{\phi}) (-\Delta_{\phi}+D-\Delta_F) \Big),
\end{align}
and
\begin{align}\label{ONppconstant2}
\mathcal{X}_2(\Delta_{\phi},\Delta_F,D)=\,&
\frac{\pi^{3D}\, \Gamma^2\big(\frac{\Delta_F}{2}\big)\, \Gamma\big(\Delta_F+1\big)\, \Gamma^3\big(\frac{D}{2}-\Delta_F\big)\, \Gamma\big(\Delta_F -\frac{D}{2}+1\big)}{2 D\Delta_F^2\, \Gamma\big(\frac{D}{2}\big)\, \Gamma\big(\frac{D}{2}-\Delta_F+1\big)\, \Gamma^2\big(\frac{1}{2}(D-\Delta_F)\big)\, \Gamma\big(D-\Delta_F+1\big)}
\nn\\
&\times \frac{\Gamma^2\big(\frac{D}{2}-\Delta_{\phi}\big)\, \Gamma\big(\Delta_{\phi}-\frac{\Delta_F}{2}\big)\,
\Gamma\big(\Delta_{\phi}-\frac{D}{2}+\frac{\Delta_F}{2}\big)}
{\Gamma^2\big(\Delta_{\phi}\big)\, \Gamma^3\big(\Delta_F\big)\, \Gamma\big(\frac{D}{2} + \frac{\Delta_F}{2} - \Delta_{\phi}\big) \Gamma\big(D-\Delta_{\phi}-\frac{\Delta_F}{2}\big)}
\nn\\
&\times \Big(
D^2 \Delta_F\, \psi\big(\tfrac{D}{2} - \Delta_{\phi} + \tfrac{\Delta_F}{2}\big)
- D^2 \Delta_F\, \psi\big(\Delta_{\phi}-\tfrac{D}{2}+\tfrac{\Delta_F}{2}\big)
\nn\\
&\qquad +D \Delta_F (\Delta_F-D)\, \psi\big(D-\Delta_{\phi}-\tfrac{\Delta_F}{2}\big)
+D \Delta_F^2\, \psi\big(\Delta_{\phi}-\tfrac{D}{2}+\tfrac{\Delta_F}{2}\big)
\nn\\
&\qquad +D \Delta_F (D-\Delta_F)\, \psi\big(\Delta_{\phi}-\tfrac{\Delta_F}{2}\big)
-D \Delta_F^2\, \psi\big(\tfrac{D}{2} - \Delta_{\phi} + \tfrac{\Delta_F}{2}\big)
\nn\\
&\qquad -2 D^2 + 4 D \Delta_{\phi}+4 D \Delta_F-8 \Delta_{\phi} \Delta_F
\Big).
\end{align}
The two functions $\mathcal{X}_1(\Delta_{\phi},\Delta_F,D)$ and $\mathcal{X}_2(\Delta_{\phi},\Delta_F,D)$ are introduced in \eqref{On-booteqns-I-facctor}.

By the way, let us also list the function $\mathcal{I}(\Delta_{\phi},\Delta_F,D)$ introduced in \eqref{On-booteqns-I-facctor}, corresponding to the vertex bootstrap equation \eqref{boot-F},
\begin{align}
\mathcal{I}(\Delta_{\phi},\Delta_F,D) =\,& 
\kappa(D-\Delta_F,D-\Delta_F,2 \Delta_F)\, \kappa(D-\Delta_F,D-2\Delta_{\phi}+\Delta_F,2 \Delta_{\phi })
\nn\\&\times \kappa(2\Delta_{\phi}, D-\Delta_F, D - 2\Delta_{\phi} + \Delta_F)\,\hat{\mathcal{I}}(\Delta_{\phi},\Delta_F,D),
\end{align}
where $\hat{\mathcal{I}}(\Delta_{\phi},\Delta_F,D)$ is give by 
\begin{align}\label{ONvertexconstant}
\hat{\mathcal{I}}(\Delta_{\phi},\Delta_F,D) =\,&
\int_{-i \infty}^{i\infty} ds_1 ds_2 dt_1 dt_2 
\frac{\pi^{D}\,\Gamma \big(\tfrac{D}{3}-\tfrac{\Delta_F}{2}+s_2\big)\,
\Gamma\big(\tfrac{D}{4} -\tfrac{\Delta_{\phi}}{2} + \tfrac{1}{2}s_2 + t_2\big)}
{\Gamma^2\big(\tfrac{\Delta_F}{2}\big)\, \Gamma^2\big(D-\Delta_{\phi}-\tfrac{\Delta_F}{2}\big)\,
\Gamma\big(\tfrac{D}{2} -\tfrac{\Delta_{\phi}}{4} + \tfrac{1}{2}s_1 + t_1\big)}
\nn\\
&\times \frac{\Gamma \big(\tfrac{D}{4} - \tfrac{\Delta_{\phi}}{4} -\tfrac{1}{2}s_1 - \tfrac{1}{2}s_2 - t_1 - t_2\big)\,
\Gamma \big(\tfrac{D}{2} -\tfrac{\Delta_{\phi}}{4} - \tfrac{\Delta_F}{2} + \tfrac{1}{2}s_1 + t_1\big)}
{\Gamma \big(\tfrac{D}{4} - \tfrac{\Delta_{\phi}}{4}+ \tfrac{1}{2}\Delta_F- \tfrac{1}{2}s_1- \tfrac{1}{2}s_2 - t_1 - t_2\big)\,
\Gamma \big(\tfrac{\Delta_{\phi}}{2}+\tfrac{D}{3}-\tfrac{\Delta_F}{2}+s_1\big)}
\nn\\
&\times \frac{\Gamma \big(\tfrac{\Delta_{\phi}}{4} - \tfrac{D}{3} + \tfrac{\Delta_F}{2} - \tfrac{1}{2}s_1 + t_1\big)\,
\Gamma \big(\tfrac{\Delta_{\phi}}{4}+\tfrac{D}{4} - \tfrac{\Delta_F}{2}+ \tfrac{1}{2}s_1 + \tfrac{1}{2}s_2 + t_1 + t_2\big)}
{\Gamma\big(\Delta_{\phi}-\tfrac{\Delta_F}{2}\big)\,
\Gamma\big(\tfrac{\Delta_{\phi}}{2} + \tfrac{D}{4}- \tfrac{\Delta_F}{2} + \tfrac{1}{2}s_2+ t_2\big)\,
\Gamma\big(\tfrac{\Delta_{\phi}}{2} - \tfrac{D}{6} + \tfrac{\Delta_F}{2} - s_1 - s_2\big)}
\nn\\
&\times
\Gamma \big(\tfrac{D}{4} -  \tfrac{\Delta_{\phi}}{2}+ \tfrac{\Delta_F}{2} - \tfrac{1}{2}s_2- t_2\big)
\Gamma \big(\tfrac{D}{3} -\tfrac{\Delta_{\phi}}{2} +s_1\big)
\Gamma\big(\tfrac{D}{3} - \tfrac{\Delta_{\phi}}{2} - s_1 - s_2\big)
\nn\\
& \times 
\Gamma \big(\tfrac{\Delta_{\phi}}{2}-\tfrac{D}{6}+s_1\big)\,
\Gamma\big(\tfrac{\Delta_{\phi}}{4} - \tfrac{D}{12} + \tfrac{1}{2} s_1 + \tfrac{1}{2}s_2 - t_1 - t_2\big).
\end{align}
In $D=3$, we find that the contours such that all the $\Gamma$'s in the numerator of the integrand having positive real parts only if $(\Delta_{\phi},\Delta_F)$ is chosen to be inside the diamond-like region given by the four points 
\be
(\Delta_{\phi},\Delta_F)=\{(0.75,1.5),(1.5,3), (2.25,1.5), (1.5,0)\}.
\ee
The $O(N)$ vector models, however, lives beyond this region.
We have analytically continued the integral in \eqref{ONvertexconstant} to the physical regions in $D=3$ using \texttt{MB.m}, and then numerically computed the resulting MB integrals with \texttt{Cuba}.
See Section \ref{ONmodels} for a detailed analysis about our results.

\bibliographystyle{JHEP}
\bibliography{CFT-Bib.bib}

\end{document}